\documentclass[preprint]{aastex63}
\pdfoutput=1 %for arXiv submission
\usepackage{amsmath,amstext}
\usepackage{nicefrac}
\usepackage{wasysym} 						
\usepackage{verbatim}
\usepackage{soul}
\usepackage[T1]{fontenc}
\usepackage{apjfonts} 
\usepackage{hyperref}
\usepackage{multirow}
\usepackage{natbib}

\newcommand{\etal}{~et~al. } 
\shorttitle{The tSZ Effect from Massive 0.5 $\leq$ z $\leq$ 1.5 Galaxies}
\shortauthors{Meinke \etal}

\begin{document}

%%%%%%%%%%%%%%%%%%%%%%%%%%%%%%%%%%%%%%%%%%%%%%%%%%%%%%%%%%%%%%%%
%
\title{The Thermal Sunyaev-Zel'dovich Effect from Massive, Quiescent 0.5 $\leq$ z $\leq$ 1.5 Galaxies}
\author{Jeremy Meinke}
\affiliation{Department of Physics, Arizona State University, P.O. Box 871504, Tempe, AZ 85287, USA}

\author{Kathrin B\"ockmann}
\affiliation{Universitat Hamburg, Hamburger Sternwarte, Gojenbergsweg 112, 21029, Hamburg, Germany}

\author{Seth Cohen}
\affiliation{School of Earth and Space Exploration, Arizona State University, P.O. Box 876004, Tempe, AZ 85287, USA}

\author{Philip Mauskopf}
\affiliation{Department of Physics, Arizona State University, P.O. Box 871504, Tempe, AZ 85287, USA}
\affiliation{School of Earth and Space Exploration, Arizona State University, P.O. Box 876004, Tempe, AZ 85287, USA}

\author{Evan Scannapieco}
\affiliation{School of Earth and Space Exploration, Arizona State University, P.O. Box 876004, Tempe, AZ 85287, USA}

\author{Richard Sarmento}
\affiliation{United States Naval Academy,  121 Blake Road, Annapolis, MD, 21402, USA}

\author{Emily Lunde}
\affiliation{School of Earth and Space Exploration, Arizona State University, P.O. Box 876004, Tempe, AZ 85287, USA}

\author{J'Neil Cottle}
\affiliation{School of Earth and Space Exploration, Arizona State University, P.O. Box 876004, Tempe, AZ 85287, USA}

%\author{Jeremy Meinke$^{1}$, Kathrin B\"ockmann$^2$, Seth Cohen$^{3}$, Philip Mauskopf$^{1,3},$ Evan Scannapieco$^3$, Emily Lunde$^3$, \& J'Neil Cottle$^3$}
%\altaffiltext{1}{Department of Physics, Arizona State University, P.O. Box 871504, Tempe, AZ 85287, USA}
%\altaffiltext{2}{Universitat Hamburg, Hamburger Sternwarte, Gojebergsweg 112, 21029, Hamburg, Germany}
%\altaffiltext{3}{School of Earth and Space Exploration, Arizona State University, P.O. Box 876004, Tempe, AZ 85287, USA}
%\date{}  is the most important in the case of the 

%%%%%%%%%%%%%%%%%%%%%%%%%%%%%%%%%%%%%%%%%%%%%%%%%%%%%%%%%%%%%%%%

% ABSTRACT
\begin{abstract} 

We use combined South Pole Telescope (SPT)+Planck temperature maps to analyze the circumgalactic medium (CGM) encompassing 138,235 massive, quiescent 0.5 $\leq$ z $\leq$ 1.5 galaxies selected from data from the Dark Energy Survey (DES) and Wide-Field Infrared Survey Explorer (WISE).  Images centered on these galaxies were cut from the 1.85 arcmin resolution maps with frequency bands at 95, 150, and 220 GHz. The images were stacked, filtered, and fit with a gray-body dust model to isolate the thermal Sunyaev-Zel'dovich (tSZ) signal, which is proportional to the total energy contained in the CGM of the galaxies.  We separate these $M_{\star} = 10^{10.9} M_\odot$ - $10^{12} M_\odot$ galaxies into 0.1 dex stellar mass bins, detecting tSZ per bin up to $5.6\sigma$ and a total signal-to-noise ratio of $10.1\sigma$.  We also detect dust with an overall signal-to-noise ratio of $9.8\sigma$, which overwhelms the tSZ at 150GHz more than in other lower-redshift studies.  We correct for the $0.16$ dex uncertainty in the stellar mass measurements by parameter fitting for an unconvolved power-law energy-mass relation, $E_{\rm therm} = E_{\rm therm,peak} \left(M_\star/M_{\star,{\rm peak}} \right)^\alpha$, with the peak stellar mass distribution of our selected galaxies defined as $M_{\star,{\rm peak}}= 2.3 \times 10^{11} M_\odot$. This yields an $E_{\rm therm,peak}= 5.98_{-1.00}^{+1.02} \times 10^{60}$ erg and $\alpha=3.77_{-0.74}^{+0.60}$.  These are consistent with $z \approx 0$ observations and within the limits of moderate models of active galactic nuclei (AGN) feedback.  We also compute the radial profile of our full sample, which is similar to that recently measured at lower-redshift by \cite{Schaan2020}.

\end{abstract}

%\tableofcontents
%\newpage

%\titleformat{\section}[runin]
%  {\normalfont\Large\bfseries}{\thesection}{1em}{}
%\titleformat{\subsection}[runin]
%  {\normalfont\large\bfseries}{\thesubsection}{1em}{}
%  \titleformat{\subsubsection}[runin]
%  {\normalfont\normalsize\bfseries}{\thesubsubsection}{1em}{}

\pagenumbering{arabic}
\pagestyle{plain}


\section{Introduction}

We have yet to understand the processes that shaped the history of the most massive galaxies in the universe.  Each such galaxy is made up of a bulge of old, red stars surrounding a massive black hole, and the two are strongly connected. To be consistent with cosmological constraints, the lack of young stars in these galaxies requires active galactic nuclei (AGN) to have exerted significant feedback on their environment \citep{Silk98,Granato2004,Scannapieco2004,Croton2006,Bower2006}.  Without feedback from AGN, progressively more massive galaxies would form at later times \citep{Rees1977,White1991}, in direct contrast with observations, which show that the typical mass of star-forming galaxies has been decreasing for the last $\approx$ 10 Gyrs \citep{Cowie1996,Treu2005,Drory2008}. 

Yet, the details of AGN feedback remain uncertain. Two types of feedback models have been proposed.  In `quasar mode' feedback, the circumgalactic medium (CGM) is impacted by a powerful outburst that occurs when the supermassive black hole is accreting most rapidly.   In this case, CGM is heated to a high enough temperature and entropy that the gas cooling time is much longer than the Hubble time, suppressing further star formation until today.  Such models are supported by observations of high-velocity flows of ionized gas associated with the black holes accreting near the Eddington rate
 \citep{Harrison2014, Greene2014,Lansbury2018,Miller2020}, but the mass and energy flux from such quasar is difficult to constrain due to uncertain estimates of the distance of the outflowing material from the central source \citep{Wampler1995,deKool2001,Chartas2007,Feruglio2010,Dunn2010,Veilleux2013,Chamberlain2015}.  
  
 In `radio mode' feedback, on the other hand, cooling material is more gradually prevented from forming stars by jets of relativistic particles that arise during periods of lower accretion rates.  In this case, the CGM is maintained at a roughly constant temperature and entropy, as low levels of gas cooling are continually balanced by energy input from the relativistic jets.
Such models are supported by AGN observations of lower power jets of relativistic plasma \citep{Fabian2012}.  These couple efficiently to the volume-filling hot atmospheres of galaxies clusters \citep{McNamara2000,Churazov2001,McNamara2016}, but may or may not play a significant role in balancing cooling in less massive gravitational potentials \citep{Werner2019}.
 
 One of the most promising methods for distinguishing between these models is by looking at anisotropies in the cosmic microwave background (CMB) photons passing through hot, ionized gas. If the gas is sufficiently heated, it will impose observable redshift-independent fluctuations in the CMB known as the thermal Sunyaev-Zel'dovich (tSZ) effect \citep{Sunyaev1972}.  The resulting CMB anisotropy has a distinctive frequency dependence, which causes a deficit of photons below $\nu_{\text{null}} = 217.6 \, \text{GHz}$ and an excess of photons above $\nu_{\text{null}}$. The change in CMB temperature $\Delta T$ as a function of frequency due to the (non-relativistic) tSZ effect is given by
\begin{equation} 
\frac{\Delta T}{T_{\text{CMB}}} = y \left( x \frac{e^x + 1}{e^x - 1} - 4 \right),
\label{eq:DeltaT}
\end{equation}
where the dimensionless Compton-$y$ parameter is defined as
\begin{equation}
 y \equiv \int dl \, \sigma_T \frac{n_e k \left( T_e - T_{\rm CMB} \right)}{m_e c^2}, 
 \label{eq:y}
\end{equation}
 where $\sigma_T$ is the Thomson cross section, $k$ is the Boltzmann constant, $m_e$ is the electron mass,
 $c$ is the speed of light, $n_e$ is the electron number density, $T_e$ is the electron temperature, $T_{\text{CMB}}$ is the CMB temperature (we use $T_{\text{CMB}} = 2.725$ K), the integral is performed over the line-of-sight distance $l$, and the dimensionless frequency $x$ is given by $x \equiv h \nu / k T_{\text{CMB}} = \nu / 56.81 \,\text{GHz}$, where $h$ is the Planck constant. 
 
 As the Compton-$y$ parameter is proportional to both $n_e$ and $T,$ it provides a measure of the total pressure along the line-of-sight.   Therefore by integrating the tSZ signal over a patch of sky, we can obtain the volume integral of the pressure, and calculate the total  thermal energy $E_{\rm therm}$ in the CGM associated with a source \citep[e.g.][]{Scannapieco2008,Mroczkowski2019}.
As detailed in \cite{Spacek2016}, this gives 
\begin{equation}
E_{\rm therm} = 2.9 \times 10^{60}  \left(\frac{l_{\rm ang}}{\text{Gpc}}\right)^2 
\frac{\int \Delta y(\vec{\theta}) d\vec{\theta}}{\text{$10^{-6}$ arcmin$^2$}},
\label{eq:EthrmT}
\end{equation}
where throughout this work, we adopt a $\Lambda$CDM cosmological model with parameters \citep[from][]{PlanckVI}, $h=0.68$, $\Omega_0$ = 0.31, $\Omega_\Lambda$ = 0.69, and $\Omega_b = 0.049$, where $h$ is the Hubble constant in units of 100 km s$^{-1}$ Mpc$^{-1}$, and $\Omega_0$, $\Omega_\Lambda$, and $\Omega_b$ are the total matter, vacuum, and baryonic densities, respectively, in units of the critical density. 

This relationship means that improvements in the sensitivity and angular resolution of tSZ measurements translate directly into improvements in constraints on thermal energy.   Thus the cosmic structures with higher gas thermal energies, galaxy clusters, are most easily detected by tSZ measurements, and indeed, measurements of the tSZ effect over the last decade have been focused on detecting and characterizing these structures \citep[e.g.][]{PlanckVIII,Reichardt2013,PlanckXXIV,Hilton2018}.   

On the other hand, pushing to lower mass halos has proven to be much more challenging. While in one case, evidence of a tSZ decrement caused by outflowing gas associated with a single luminous quasar was found in ALMA measurements \citep{Lacy2019}, most such constraints have involved averaging over many objects.  In this regard, \citet{Chatterjee2010} used data from the Wilkinson Microwave Anisotropy Probe and Sloan Digital Sky Survey (SDSS) around both quasars and galaxies to find a tentative $\approx 2\sigma$ tSZ signal suggesting AGN feedback; \citet{Hand2011} used data from SDSS and the Atacama Cosmology Telescope (ACT)  to find a $\approx 1\sigma-3\sigma$  tSZ signal around galaxies; \citet{Gralla2014} used the ACT to find a $\approx 5\sigma$  tSZ signal around AGNs;  \citet{Ruan2015} used SDSS and Planck to find $\approx 3.5\sigma-5.0\sigma$ tSZ signals around both quasars and galaxies; \citet{Crichton2016} used SDSS and ACT to find a $3\sigma-4\sigma$ SZ signal around quasars; \citet{Hojjati2016} used data from Planck and the Red Cluster Sequence Lensing Survey to find a $\approx 7\sigma$ tSZ signal suggestive of AGN feedback; and \citep{Hall2019} used ACT, 
Herschel, and the Very Large Array data to measure the tSZ effect around $\approx 100,000$ optically selected quasars, finding a $3.8 \sigma$ signal that provided a joint constraint on AGN feedback and mass of the quasar host halos at $z \gtrsim 2.$

Recent measurements have also been made around massive galaxies.   At $z \lesssim 0.5$, \citet{Greco2015} used SDSS and Planck data to compute the average tSZ signal  from a range of over 100,000 `locally brightest galaxies' (LBGs).  This sample was large enough to derive constraints on $E_{\rm therm}$  as a function of galaxy stellar mass $M_\star$ for objects with $M_\star \gtrsim 2 \times 10^{11} M_\odot.$ More recently, \citet{Schaan2020} and \citet{Amodeo2020} combined microwave maps from ACT DR5  and Planck with in the galaxy catalogs from the Baryon Oscillation Spectroscopic Survey (BOSS), to study the gas associated with these galaxy groups. They measured the tSZ signal at $\approx 10 \sigma$ along with a weaker detection of the kinetic Sunyaev-Zel'dovich effect \citep{Sunyaev1980}, which constrains the gas density profile.  They were able to compare these results to cosmological simulations \citep{Battaglia2010,Springel2018} to find that the feedback employed in these models was insufficient to account for the gas heating observed at $\approx$ Mpc scales.

At redshifts $0.5 \lesssim z \lesssim 1.5$ the SZ signal from massive quiescent galaxies was studied by \citet{Spacek2016,Spacek2017}. These are precisely the objects for which AGN feedback is thought to quench star formation and thus where a significant excess tSZ signal is expected \citep[e.g.][]{Scannapieco2008}. To obtain this faint signal, \cite{Spacek2016} performed a stacking analysis using the VISTA Hemisphere Survey and Blanco Cosmology Survey data overlapping with 43 deg$^2$ at 150 and 220 GHz from the 2011 South Pole Telescope (SPT) data release, finding a $\approx 2-3 \sigma$ signal hinting at non-gravitational heating. \cite{Spacek2017} used SDSS and the Wide-Field Infrared Survey Explorer (WISE) data overlapping with 312 deg$^2$ of 2008/2009 ACT data at 148 and 220 GHz, finding a marginal detection that was consistent with gravitational-only heating models. 

Here we build on these measurements by making use of data from WISE, the Dark Energy Survey, and the 2500 square-degree survey of the southern sky taken by the SPT, which includes measurements at 95, 150, and 220 GHz.  The increase in frequency and sky coverage allows us to obtain a $10.1\sigma$ detection of the tSZ effect around $z \approx 1$ galaxies.  This lets us move from the marginal detections and upper limits presented in our previous work to measurements that can be applied to future simulations to strongly distinguish between feedback models.  

The structure of this paper is as follows: In \S\ref{sec:data} we describe the data sets used for our analysis and contrast that with our previous work.  In \S\ref{sec:sample_selection} we describe our galaxy selection procedure, and the overall properties of the sample of massive, moderate-redshift, quiescent galaxies we use for the stacking analysis presented in \S\ref{sec:StackingFiltering}.  In \S\ref{sec:Results}, we contrast our measurements with the work from other groups as well as with simple feedback models.  Conclusions are given in \S\ref{sec:Discussion}.

\section{Data}
\label{sec:data}
 
For our analysis, we use three public datasets, two to detect and select galaxies, and one to make our tSZ measurements.  As discussed in \S\ref{sec:sample_selection}, selecting and carrying out photometric fitting of passive galaxies at $0.5 \lesssim z \lesssim 1.5$ requires data that spans optical, near infrared, and mid-infrared wavelengths.  Thus we make use of optical and near-infrared data from DES data release 1 \citep{Abbott2018},  which are already matched to AllWISE data spanning 3-25 $\mu m$ \citep{Schlafly2019}.  For detecting the tSZ effect, we use millimeter-wave observations from the SPT-SZ survey \citep{Bocquet2019}. The three datasets, which overlap over an area of $\approx$ 2500 deg$^{2},$ are described in more detail below.

\subsection{DES}
DES DR1 is based on optical and near-infrared imaging from 345 nights between August 2013 to February 2016 by the Dark Energy Camera mounted on the 4-m Blanco telescope at Cerro Tololo Inter-American Observatory in Chile. The data covers $\approx$ 5000 deg$^2$ of the southern Galactic cap in five photometric bands: grizY. These five bands have point-spread functions of g = 1.12, r = 0.96, i = 0.88, z = 0.84, and Y = 0.90 arcsec FWHM \citep{Abbott2018}.   The survey has exposure times of 90s for griz and 45s for Y band, yielding a typical single-epoch PSF depth at S/N = 10 for g $\lesssim$ 23.57, r $\lesssim$ 23.34, i $\lesssim$ 22.78, z $\lesssim$ 22.10 and Y $\lesssim$ 20.69 mag \citep{Abbott2018}. Here and below, all magnitudes are quoted in the AB system \citep[i.e.][]{Oke1983}.

\subsection{WISE}
The AllWISE catalog is based on the Wide-field Infrared Survey Explorer (WISE) NASA Earth orbit mission \citep{wise10,neowise11}. In 2010 WISE carried out an all-sky survey of the sky in bands W1, W2, W3 and W4, centered at 3.4, 4.6, 12 and 22 $\mu$m, respectively \citep{Schlafly2019}. The 40 cm diameter infrared telescope was equipped with four 1024x1024 pixel focal plane detector arrays cooled by a dual-stage solid hydrogen cryostat.  The whole sky was surveyed 1.2 times in all four bands at a full sensitivity. After the hydrogen ice in the outer cryogen tank evaporated, WISE surveyed an additional third of the sky in three bands, with the W1 and W2 detectors operating at near full sensitivity while the W3 focal plane operated at a lower sensitivity \citep{neowise11}.

AllWISE uses the work of the WISE mission by combining data from the cryogenic and post-cryogenic survey, which yields a deeper coverage in the W1 and W2. The added sensitivity of AllWISE extends the limit of detection of luminous distant galaxies because their apparent brightness at 4.6 $\mu$m (W2) no longer declines significantly with increasing redshift. The increased sensitivity yields better detection of those galaxies for redshift $z > 1$. This is crucial for our galaxy detection and selection because we are especially looking for luminous distant galaxies at a redshift $z > 1$.

\subsection{SPT-SZ}

The SPT-SZ survey \citep{Chown2018} covers 2500 deg$^2$ of the southern sky between 2007 to 2011 in three different frequencies: 95 GHz and 150 GHz, which lie on either side of the maximum tSZ intensity decrement ($\sim128$ GHz), and 220 GHz, which is very near $\nu_{\rm null}$ = 217.6 GHz  where there is no change in the CMB signal due to the tSZ effect.  The South Pole Telescope (SPT) is a 10 m telescope located within 1 km of the geographical South Pole and consists of a 960-element bolometer array of superconducting transition edge sensors. The maps used in this analysis are publicly available\footnote{\url{https://lambda.gsfc.nasa.gov/product/spt/index.cfm}} combined maps of the SPT with data from the all-sky Planck satellite (with similar bands at 100, 143, and 217 GHz).  Each combined map has a beam resolution of 1.85 arcmin, and is provided in a HEALPix (Hierarchical Equal Area isoLatitude Pixelation) format with $N_{\rm side}=8192$ \citep{Chown2018}.  For comparison, the previous analysis of  \cite{Spacek2016} relied on the 2011 SPT data release covering a limited 95 deg$^2$ at only 150 and 220 GHz \citep{Schaffer2011}.  The addition of the 95 GHz band allows for better extraction of the tSZ component, while the larger available field increases the galaxy sample size for reduced noise.

\begin{figure*}[t]
\centering\includegraphics[height=6.5cm]{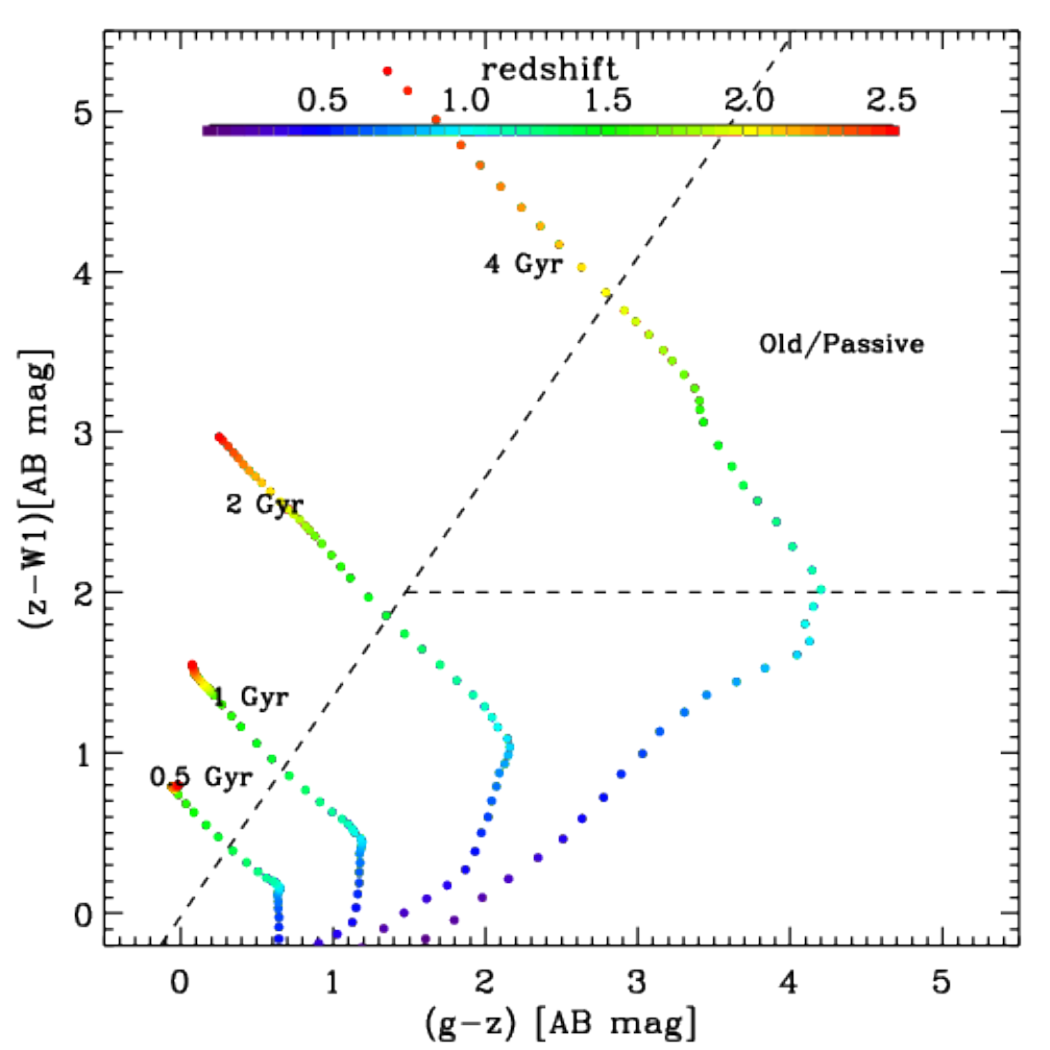}
\caption{\small Color-color plot showing selection region for passive galaxies. This is slightly modified from \cite{Spacek2017}. These are BC03 models with age as indicated. This pre-selection is only used to query the DES database, with the final selection based on SED-fit parameters.}
\label{fig:colcol}
\end{figure*}

\section{Defining the Galaxy Sample}\label{sec:sample_selection}

\begin{figure*}[ht]
\centering\includegraphics[height=6.5cm]{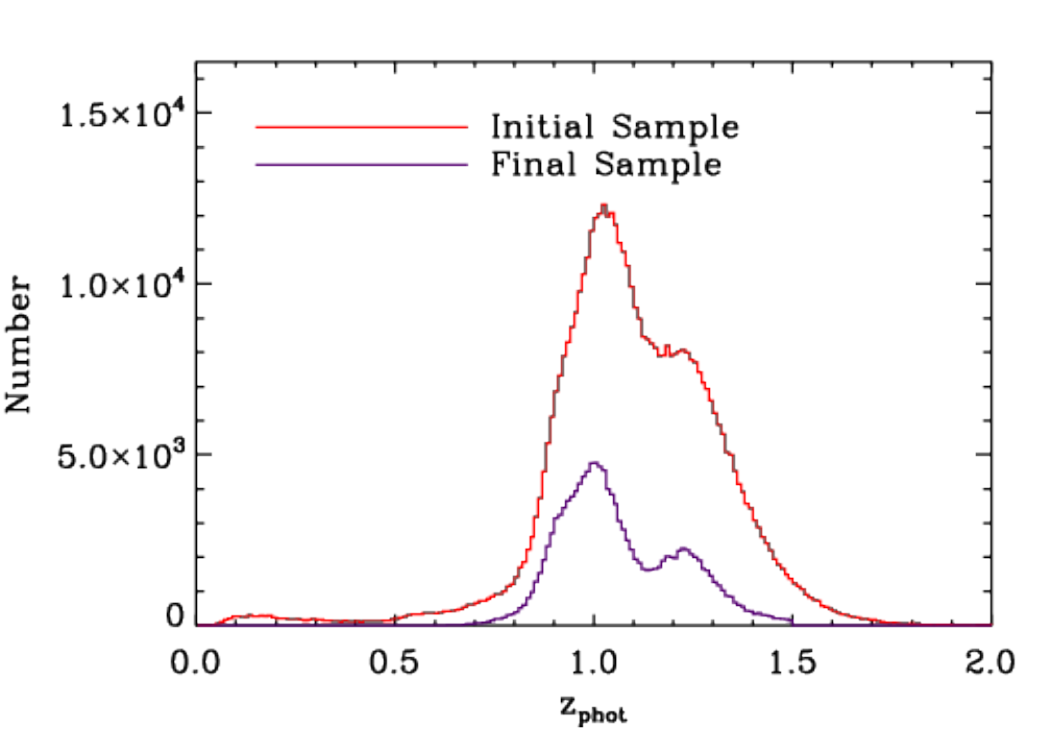}
\caption{\small Six-band photometric redshift distribution
of the initial color-selected sample (red). As expected, the majority of the sample is at $z>1$. The final sample after selecting based on goodness-of-fit, redshift, age, and SSFR is shown in purple.}
\label{fig:zhist}
\end{figure*}

\begin{figure*}[ht]
\centering\includegraphics[height=9.0cm]{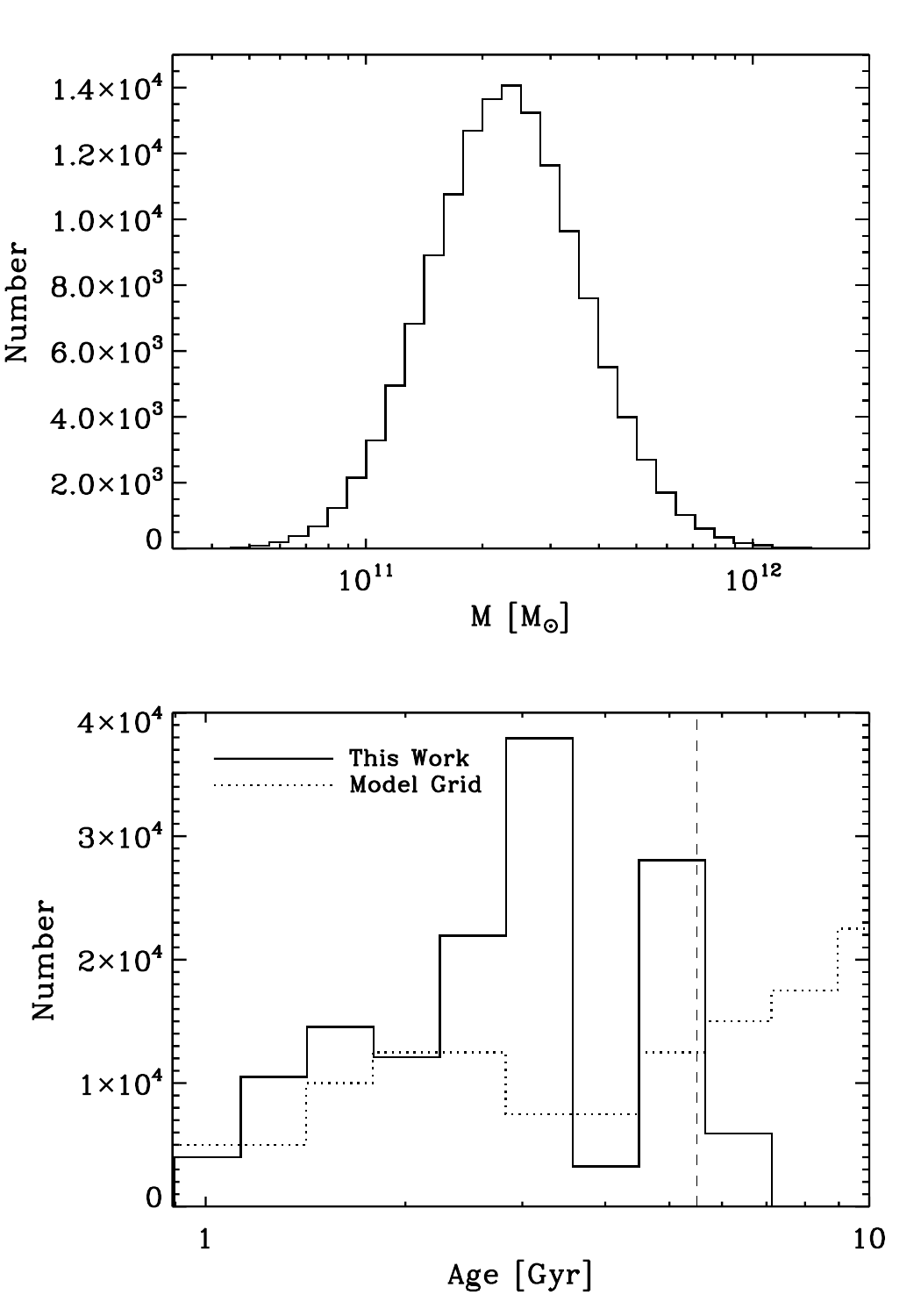}
\caption{\small {\em Top:} Stellar mass distribution of the sample after selecting based on SED parameters. {\em Bottom:} Age histogram of the sample. Note that, at a given redshift, the ages are restricted to be younger than the age of the universe at that redshift. The black dashed line shows the age of the universe at $z\!=\!1.1$. The dotted line shows the scaled distribution of all available \cite{bc03} models on the grid.}
\label{fig:masshist}
\end{figure*}

\subsection{Selection}

We carried out our initial galaxy selection  using the DES database server at NOAO, called NOAO-Lab.  In order to start with a manageable sample, we applied a cut in color-color space designed to select old galaxies with low star-formation rates at approximately $1.0 \leq z \leq 1.5$ in the initial database query, as shown in Fig.~\ref{fig:colcol}. We used mag\_auto from the DES in {\it grizy} bands, along with {\it W1} and {\it W2} PSF-magnitudes (converted to AB-system) from AllWISE \citep{wise10,neowise11} joined to the main DES table. The bands and color-selection used here are slightly different than \cite{Spacek2017} used in SDSS Stripe 82.

The NOAO Data lab allows direct queries in SQL via Jupyter notebook on their server. The lines we used to make the color selection were {\tt ((mag\_auto\_z\_dered-(w1mpro+2.699))} {\tt <= \\ (1.37*mag\_auto\_g\_dered-1.37*mag\_auto\_z\_dered-0.02))} and \\ {\tt ((mag\_auto\_z\_dered-(w1mpro+2.699) )>=2.0)}.

\subsection{Photometric Fitting}

\begin{table*}[t]
	\begin{center}
		\hspace{-1.0in}
		\resizebox{14cm}{!}{
			\begin{tabular}{|c|c|c|c|c|c|c|c|c|c|}
				\hline
				\multirow{2}{*}{$\text{log}_{10}(M_\star/M_\odot)$ Bin} & \multirow{2}{*}{$N$} & \multirow{2}{*}{$\overline{z}$} & \multirow{2}{*}{$\widetilde{z}$} & \multirow{2}{*}{$\text{log}_{10}(\overline{M}/M_\odot)$} & \multirow{2}{*}{$\text{log}_{10}(\widetilde{M}/M_\odot)$} & $\overline{Age}$ & $\overline{l_{ang}^2}$ & $\widetilde{l\:}_{ang}^2$ \\
				& & & & & & [Gyr] & [$\text{Gpc}^2$] & [$\text{Gpc}^2$] \\
				\hline $10.9-11.0$ & 3376 & 0.94 & 0.95 & 10.96 & 10.96 & 1.69 & 2.76 & 2.78 \\
				$11.0-11.1$ & 8241 & 0.97 & 0.96 & 11.06 & 11.06 & 2.05 & 2.79 & 2.81 \\
				$11.1-11.2$ & 15738 & 1.00 & 0.98 & 11.16 & 11.16 & 2.47 & 2.83 & 2.84 \\
				$11.2-11.3$ & 23448 & 1.03 & 1.01 & 11.25 & 11.25 & 2.86 & 2.87 & 2.87 \\
				$11.3-11.4$ & 27723 & 1.07 & 1.04 & 11.35 & 11.35 & 3.19 & 2.91 & 2.91 \\
				$11.4-11.5$ & 24877 & 1.09 & 1.07 & 11.45 & 11.45 & 3.52 & 2.93 & 2.94 \\
				$11.5-11.6$ & 17246 & 1.11 & 1.09 & 11.55 & 11.54 & 3.71 & 2.95 & 2.96 \\
				$11.6-11.7$ & 9501 & 1.13 & 1.13 & 11.65 & 11.64 & 3.88 & 2.97 & 3.00 \\
				$11.7-11.8$ & 4396 & 1.15 & 1.16 & 11.74 & 11.74 & 4.01 & 2.99 & 3.03 \\
				$11.8-11.9$ & 1625 & 1.17 & 1.21 & 11.84 & 11.84 & 4.14 & 3.01 & 3.06 \\
				$11.9-12.0$ & 506 & 1.21 & 1.25 & 11.94 & 11.93 & 4.13 & 3.04 & 3.09 \\
				\hline
			\end{tabular}
		}
	\end{center}
	\caption{\small Statistics of 0.1-wide dex stellar mass bins from $\text{log}_{10}(M_\star/M_\odot)=10.9-12.0$. Both mean and median are listed for redshift, mass, and angular-diameter-distance-squared.  \vspace{2mm}}
	\label{tab:meanvals}
\end{table*}

After the galaxies were selected, photometric redshifts were computed using EAZY \citep{eazy08} and the seven broad bands {\it grizyW1W2}. In calling EAZY, we used the CWW$+$KIN \citep{cww80,kinney96} templates, and did not allow for linear combinations. Since we are looking for red galaxies and have a gap in wavelength coverage between {\it y}-band and {\it W1}, we were worried that allowing combinations of templates would yield unreliable redshifts, where e.g., a red template was fit to the IR-data and a blue one was fit to the optical data and they meet in the wavelength gap. The resulting redshift distribution is shown in Fig.~\ref{fig:zhist}, which clearly shows that we selected galaxies in the desired range. 

Once the redshifts were measured, we fit the spectral energy distributions (SEDs) using our own code, following the method used in \cite{Spacek2017}, to which the reader is referred for more details. Briefly, a grid of BC03 \citep{bc03} models with exponentially declining star formation rates (SFRs) was fit over a range of stellar ages, SFHs (i.e., $\tau$), and dust-extinction values ($0<A_V<4$).  Our code uses BC03 models assuming a Salpeter initial mass function (IMF), but in comparison with the literature, we convert all stellar masses to the value assuming a Chabrier IMF ($0.24$ dex offset; \citet{Santini2015}).  As in \cite{Spacek2017}, we choose as our final sample all galaxies with age$>1$ Gyr, $SSFR <0.01 {\rm Gyr}^{-1}$, $0.5<z_{\rm phot}<1.5$, and reduced $\chi^{2}<5$. 

\subsection{Removing Known Contaminants}

Before using this catalog, there are several contaminants that must be removed. We therefore remove sources from the \emph{ROSAT} Bright and Faint Source catalogs \citep[BSC and FSC;][]{voges99}. We additionally remove known clusters from \emph{ROSAT} \citep{Piffaretti2011} and \emph{Planck} \citep{Planck2015-XXVII}. Sources from the \emph{AKARI/FIS} Bright Source Catalog \citep{AKARI/FIS} and the \emph{AKARI/IRC} Point Source Catalog \citep{AKARI/IRC} were also removed along with galactic molecular clouds by cross-matching with the Planck Catalogue of Galactic Cold Clumps \citep{Planck2015-XXVIII} and compact sources from the nine-band Planck Catalog of Compact Sources \citep{Planck2013-XXVIII}. We also remove sources from the \emph{IRAS} Point Source Catalog \citet{IRAS1994}. We \emph{do not} remove the SZ sources selected from the SPT by \citep{Bleem2015}. In all cases, sources with a possible contaminant within $4\farcm0$, to match the beam of the SPT-SZ data, are flagged and those sources are removed from further consideration. This left 138,235 massive and quiescent $z\gtrsim0.5$ galaxies to include in our SZ stacks. This final sample is shown as the purple line in Fig.~\ref{fig:zhist} and stellar mass distribution shown in Fig.~\ref{fig:masshist}.  These galaxies were partitioned into 12 logarithmic stellar mass bins of width $\Delta \text{log}_{10}(M_\star/M_\odot)=0.1$ ranging from $\text{log}_{10}(M_\star/M_\odot) = $10.9 to 12.0.  The  mean and median redshift, mass, and angular-diameter-distance-squared for each of these bins are listed in Table~\ref{tab:meanvals}.  These small mass bins were chosen to more accurately fit the dust model discussed in \S\ref{sec:StackingFiltering}, which has a dependence on redshift and potential non-linearity with $\text{log}_{10}(M_\star/M_\odot)$.

\subsection{Comparison with  Previous Work}

Our current sample has several advantages to our previous work in \cite{Spacek2016}  and  \cite{Spacek2017}.    As compared to \cite{Spacek2017} , the DES data is deeper than the SDSS data, which provides for better SED-fits in addition to fainter sources. In addition, the use of the AllWISE data is superior to the Wise All-Sky Survey \citep{wise10} used in \cite{Spacek2017}, because it includes more observing time from the extended NEOWISE mission \citep{neowise11}. Again, this helps the fidelity of the SED-fits.  Second, we have slightly altered the color selection, as given above, so as to better include galaxies in the desired redshift range of $0.5 \leq z \leq 1.5$. This was successful as shown in Fig.~\ref{fig:zhist}, as compared to Fig.\ 3 of \cite{Spacek2017}. 

The third difference is the choice of photometric redshift template. Our color-selection aims to find very red galaxies as templates designed to work on the most general sets of galaxies are not ideal. This is mainly, but not only, due to large 4000 \AA\ breaks. After testing all the available templates for EAZY, we found that the best-fits were given by the empirical CWW$+$KIN templates,  because the elliptical galaxy template had a sufficiently large 4000 \AA\ break. We found that 86\% of our galaxies were best-fit by either the ''E'' or ``Sbc'' CWW templates.  By contrast, \cite{Spacek2017} used the EAZY V1.0 templates, which are based on population synthesis models. 

Finally, \cite{Spacek2016} relied on the 2011 SPT data release covering a limited 95 deg$^2$ at only 150 and 220 GHz \citep{Schaffer2011}, while the more recent data release used here both includes the  95 GHz band and covers a significantly larger area (2500 sq.\ deg.).  This is also a much larger region than the $\approx 300$ deg$^2$ of ACT data used in \cite{Spacek2017}. All four of these improvements contribute to the present study having much higher signal-to-noise measurements than our previous work.

\begin{figure*}[ht]
\centering
\includegraphics[width=1.0\linewidth,keepaspectratio]{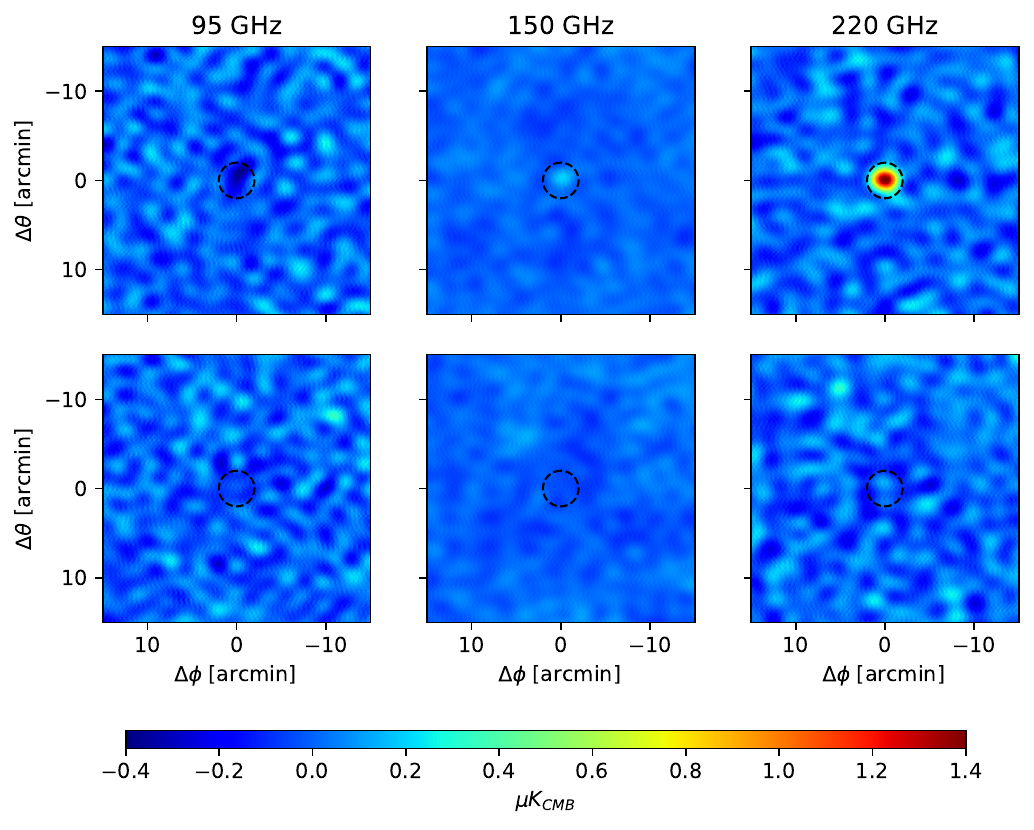}
\caption{30x30 arcmin filtered stacks at 95 (Left), 150 (Middle), and 220 (Right) GHz for  all (N=138235) galaxies (\textit{Top}), and random points (\textit{Bottom}, N=138235 out of the generated 869878). The center dashed circle in each represents the 2.0-arcmin radius top-hat aperture discussed in \S\ref{sec:Results}.}
\label{fig:AllStackZoomed}
\end{figure*}

\section{Stacking and Filtering}\label{sec:StackingFiltering}

Once the catalog of galaxies described in Table~\ref{tab:meanvals} was determined, images were taken around each galaxy location on the combined SPT-SZ maps at all three frequencies.  In the conversion from the HEALPix map format, we set the Cartesian pixel resolution of the images to 0.05 arcmin ($\approx$ 74 pixels per HEALPix pixel), so the full 60x60 arcmin images contain 1201x1201 pixels.  As the SPT region is far from the equatorial plane, the right ascension must be accurately scaled to the cosine of the declination.  We constructed averaged co-added stacked images from the individual galaxies, resulting in one stacked image per frequency per bin.

To remove large-scale CMB and dust fluctuations, we applied a 5-arcmin high-pass Gaussian filter to each averaged frequency stack.  We used an iterative Gaussian method alongside the high-pass filter to minimize central signal loss in the process.  This involved fitting the filtered stacked image center with a symmetric 2-D Gaussian,  subtracting this fit from the image prior to the high-pass filter and repeating until no further central fit could be made.  This cutoff was determined by the limit of either the Gaussian fit amplitude being less than the surrounding image noise, or the Gaussian FWHM being fit with $<10\sigma$ certainty. The Gaussian FWHM fit was also constrained to a maximum of 1.5 times the map beam resolution (2.775 arcmin) to ensure no large or runaway fits, as a $z=1$ galaxy would have an angular size $\approx1$ arcmin (for a diameter of 0.5 Mpc).  The last iterative high-pass filter was then used on the original stacked image to produce the final averaged frequency images used for signal extraction.

To account for any residual bias offset from SPT maps, galaxy selection, and filtering procedure, we also generated a set of random points  in the SPT field.  After an identical 4-arcmin cut of contaminant sources, 869,878 random points were obtained.  They were then stacked, averaged, and high-pass filtered as outlined above for the galaxies.  Measurements such as those listed in \S\ref{sec:Results} include corrections obtained from these residual bias offsets. The stacked and filtered images around galaxies are illustrated in the upper panels of  Fig.~\ref{fig:AllStackZoomed}, while the comparison stacks for the random points are illustrated in the lower panels of the figure.  There is a clearly visible signal in the galaxy stacks not seen in the random stacks.

\vspace{0.4in}

\section{Results}\label{sec:Results}

\subsection{Two Component Fitting}\label{subsec:tSZ_Dust}

To extract the central signal from each filtered frequency stack, we used a circular top-hat aperture of 2.0 arcmin radius.  This is large enough to contain the majority of the central signal while minimizing noise introduced from any other surrounding sources.  Additional apertures and sizes were also investigated as detailed in Appendix~\ref{Appendix:Aperture}. To correct for beam and filter effects, we scaled all apertures with respect to $S_{\nu,{\rm beam}}^{-1},$ where $S_{\nu,{\rm beam}}$ is the aperture signal detected from a normalized central source convolved to the beam FWHM of 1.85 arcmin, and filtered as in \S\ref{sec:StackingFiltering}.  The final 2.0 arcmin top-hat aperture used here has a scale factor of 1.03, indicating all but $3\%$ of an unresolved central source is within the aperture. Table~\ref{tab:observedvals} lists the 2.0 arcmin top-hat measurements after bias correction and scaling.

\begin{table*}[ht]
	\begin{center}
		\hspace{-0.5in}
		\def\arraystretch{1.2}
		\begin{tabular}{|c|c|c|c|c|c|}
			\hline
			\multirow{2}{*}{$\text{log}_{10}(M_\star/M_\odot)$ Bin} & \multicolumn{3}{c|}{$S_\nu=\int \Delta T_{\nu}(\vec{\theta}) d\vec{\theta}$ \hspace{0.5cm} [$\mu \text{K arcmin}^2$]} & $\int D_{220}(\vec{\theta}) d\vec{\theta}$ & $\int y(\vec{\theta}) d\vec{\theta}$ \\ 
			\cline{2-4} & 95 GHz & 150 GHz & 220 GHz & [$\mu \text{K arcmin}^2$] & [$10^{-6} \text{arcmin}^2$] \\
			\hline 
			$10.9-11.0$ & $-3.66 \pm 2.95$ & $-0.02 \pm 2.40$ & $2.40 \pm 3.80$ &  $0.34_{-0.44}^{+0.43}$ & $0.78_{-0.63}^{+0.63}$  \\
			$11.0-11.1$ & $-0.72 \pm 1.89$ & $2.20 \pm 1.54$ & $9.07 \pm 2.44$  &  $0.98_{-0.33}^{+0.31}$ & $0.47_{-0.42}^{+0.41}$ \\
			$11.1-11.2$ & $-1.04 \pm 1.37$ & $2.15 \pm 1.12$ & $3.47 \pm 1.77$ &  $0.46_{-0.22}^{+0.21}$ & $0.15_{-0.30}^{+0.30}$ \\
			$11.2-11.3$ & $-2.79 \pm 1.13$ & $-0.06 \pm 0.92$ & $5.73 \pm 1.46$ & $0.57_{-0.19}^{+0.18}$ & $0.87_{-0.25}^{+0.25}$  \\
			$11.3-11.4$ & $-1.55 \pm 1.04$ & $0.37 \pm 0.85$ & $5.37 \pm 1.34$ & $0.50_{-0.17}^{+0.15}$ & $0.59_{-0.23}^{+0.23}$ \\
			$11.4-11.5$ & $0.44 \pm 1.10$ & $3.72 \pm 0.90$ & $13.94 \pm 1.42$ & $1.26_{-0.30}^{+0.25}$ & $0.44_{-0.28}^{+0.27}$  \\
			$11.5-11.6$ & $-5.59 \pm 1.31$ & $-0.90 \pm 1.07$ & $9.28 \pm 1.69$  & $0.80_{-0.23}^{+0.21}$ & $1.69_{-0.30}^{+0.29}$  \\
			$11.6-11.7$ & $-4.16 \pm 1.76$ & $-0.07 \pm 1.44$ & $11.71 \pm 2.27$ & $0.93_{-0.29}^{+0.25}$ & $1.52_{-0.40}^{+0.39}$  \\
			$11.7-11.8$ & $-6.09 \pm 2.58$ & $-0.11 \pm 2.10$ & $17.59 \pm 3.33$ & $1.34_{-0.41}^{+0.37}$ & $2.27_{-0.59}^{+0.57}$ \\
			$11.8-11.9$ & $-2.66 \pm 4.24$ & $0.44 \pm 3.46$ & $26.14 \pm 5.47$ & $1.75_{-0.57}^{+0.53}$ & $2.23_{-0.95}^{+0.94}$  \\
			$11.9-12.0$ & $-15.52 \pm 7.60$ & $-7.53 \pm 6.19$ & $29.21 \pm 9.79$ & $1.81_{-0.84}^{+0.80}$ & $5.57_{-1.65}^{+1.65}$  \\
			\hline  
		\end{tabular}
	\end{center}
	\caption{\small 2.0 arcmin top-hat integrated temperatures determined from their respective frequency stacks (95, 150, and 220 GHz) for 0.1-wide dex stellar mass bin subsets of the galaxy catalog.  These values were extracted from the high-pass filtered stacks with the random point bias offsets subtracted, and scaled for beam correction.  The last two columns show the Dust and tSZ values obtained via component fit of eq.~(\ref{eq:SZFit}).}
	\label{tab:observedvals}
\end{table*}

\begin{figure*}[ht]
	\includegraphics[width=0.5\linewidth,keepaspectratio]{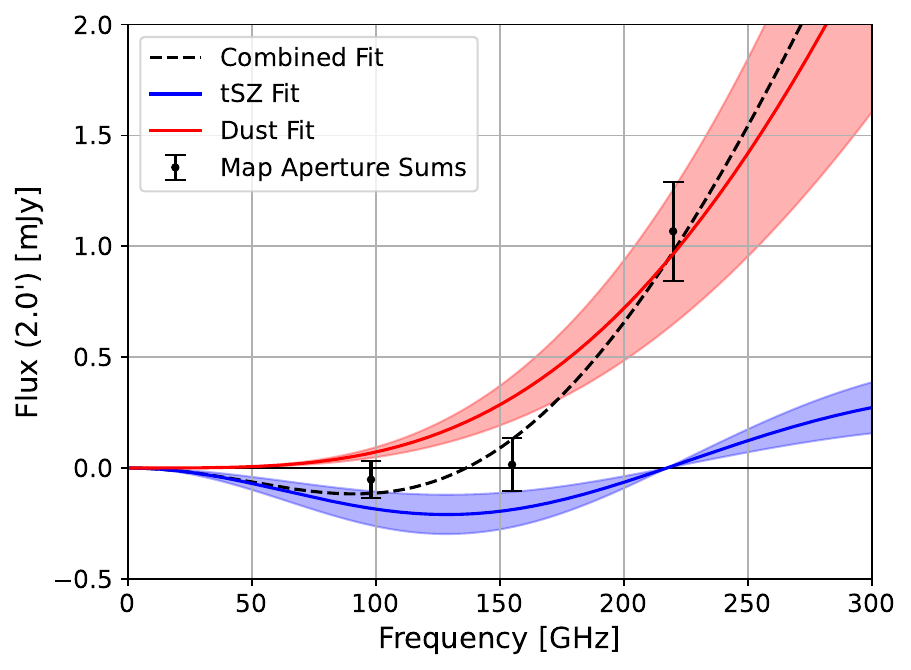}\hfill
	\includegraphics[width=0.5\linewidth,keepaspectratio]{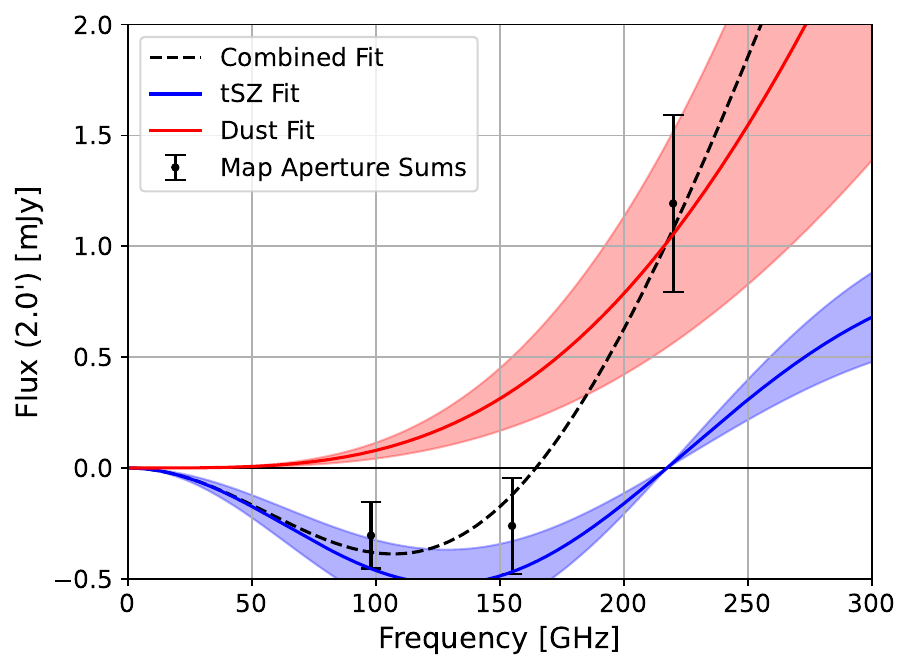}
	\caption{Intensity spectrum of our two-component fit for the two highest stellar mass bins ($\text{log}_{10}(M_\star/M_\odot)=11.8-11.9$ and $11.9-12.0$).  The shaded tSZ and dust regions represent $1\sigma$ error. The points are the 2.0 arcmin top-hat aperture values as listed in Table~\ref{tab:observedvals}, placed at each frequency band center. \textit{Left:} For $\text{log}_{10}(M_\star/M_\odot)=11.8-11.9$ bin (N=1625 galaxies). \textit{Right}: For $\text{log}_{10}(M_\star/M_\odot)=11.9-12.0$ bin (N=506 galaxies).}
	\label{fig:FitPlots}
\end{figure*}

From our aperture measurements, we used a two-component fitting model consisting of tSZ ($y$) and $z=0$ dust in the 220 GHz band ($D_{220}$),
\begin{equation}
    S_{\nu}=\int \Delta T_{\nu}(\vec{\theta}) d\vec{\theta} = 
   f_x \int y(\vec{\theta}) d\vec{\theta}+d_{\nu,220}\int D_{220}(\vec{\theta}) d\vec{\theta},
   \label{eq:SZFit}
\end{equation} 
where $f_x\equiv T_{\text{CMB}}[x \, (e^x+1)/(e^x-1)-4]$ of the tSZ signal (from eq.~\ref{eq:DeltaT}), and $d_{\nu,220}$ is the gray-body dust spectrum conversion from CMB temperature at frequency band $\nu$ to CMB temperature in the 220 GHz band at $z=0$:
\begin{equation}
d_{\nu,220}\equiv \left[\frac{(1+z) \, \nu}{220\text{GHz}}\right]^{\beta}\frac{B[(1+z) \, \nu,T_{\rm dust}]}{B(220\text{GHz},T_{\rm dust})} 
\left.\frac{dT}{dB(\nu,T)}\right\vert_{T_{\rm CMB}}
\left.\frac{dB(220\text{GHz},T)}{dT}\right\vert_{\rm T_{CMB}},
\label{eq:dust}    
\end{equation}
where $\beta$ is the dust emissivity spectral index, and $B(\nu,T)$ is the Planck distribution.  The tSZ and dust terms are integrated over SPT bandpasses extracted from \cite{Chown2018}, as the SPT+Planck maps are dominated by the SPT response for our small angular scales ($<5$ arcmin).  We use conservative values of $\beta=1.75\pm0.25$ and $T_{\rm dust}=20\pm5$K within the bounds of previous studies \citep{Draine2011,Addison2013, Planck2014-XIV}.  Unlike investigations such as \cite{Greco2015}, these values impact final results due to our higher redshift and better map resolution, that lead to an increased dust detection at lower frequencies.  We treat the dust parameters as Gaussian priors, fitting for a Gaussian distribution of $\beta$ and $T_{dust}$ with $\sigma_\beta=0.25$ and $\sigma_{T_{dust}}=5$.  The reported best fit $y$ and $D_{220}$ are the 50th percentile (median) obtained.  Error from the priors are calculated as the bounds containing $1\sigma$ ($68.27\%$), added in quadrature with the fit error.  The uncertainty in our dust parameters determined this way contributes up to $10\%$ of our final reported errors.

The integrated aperture temperatures of Table~\ref{tab:observedvals} are fit according to eq.~\ref{eq:SZFit} with their median redshifts, yielding dust and tSZ signals as listed in the final two columns in this table. The dust is kept in units of $\mu$K CMB at $z=0$ redshift and $\nu=220$GHz for easy comparison, as dust is largest in the 220 GHz band.  

We find a $0.8-4.2\sigma$ dust signal ranging between $0.34-1.81 \mu \text{K arcmin}^2$ at 220 GHz for our mass bins.  The highest dust S/N values occur in the larger mass bins, and show a trend of increasing dust with mass.  The overall signal to noise detection of dust in our data is $9.8\sigma$, and the
dust fit is heavily determined by the integrated 220 GHz values, which is also the noisiest of the three SPT bands.

For the tSZ signal, we see a negligible $0.5-3.5\sigma$ detection in the lower mass bins between $10.9\lesssim \text{log}_{10}(M_\star/M_\odot)\lesssim 11.5$.  From $\text{log}_{10}(M_\star/M_\odot)=11.5-12.0$ however, we observe a S/N up to $5.6\sigma$.   The four highest bins centered at 11.65, 11.75, 11.85, and 11.95 yield integrated $y$ values of $1.52_{-0.40}^{+0.39},$ $2.27_{-0.59}^{+0.57},$ $2.23_{-0.95}^{+0.94},$ and  $5.57_{-1.65}^{+1.65}$ $10^{-6}$  $\text{arcmin}^2,$ respectively.  The overall signal to noise ratio of our tSZ detection is $10.1\sigma$, which is a vast improvement from the $2-3 \sigma$ measurements we were able to obtain from previous data sets \citep{Spacek2016,Spacek2017}. 

Figure~\ref{fig:FitPlots} shows the intensity spectrum of the dust and tSZ signals from our two-component fit for the two highest bins using the 2.0 arcmin top-hat aperture.  The dust spectrum is near the Rayleigh-Jeans limit, but it still contributes a significant signal at the lower frequencies when compared to the fainter tSZ.  Some of the dust fit uncertainty arises from our inability to accurately determine the dust emissivity ($\beta$) and temperature ($T_{\rm dust}$), and would likely be helped by future experiments with more frequency channels.

The Compton-$y$ measurements can further be converted to thermal energy following eq.~(\ref{eq:EthrmT}), as was done in \S\ref{subsec:masscorrection}.  Numerous steps were taken to verify the stacking process.  Appendix~\ref{Appendix:Greco} details the reproduction of previous studies \citep{PlanckCollaboration2013,Greco2015} to validate the stacking code used and compare catalog selection criteria.  Fits done with different apertures and sizes are  outlined in Appendix~\ref{Appendix:Aperture}.

\begin{figure*}[t]
	\centering\includegraphics[width=0.7\linewidth, keepaspectratio]{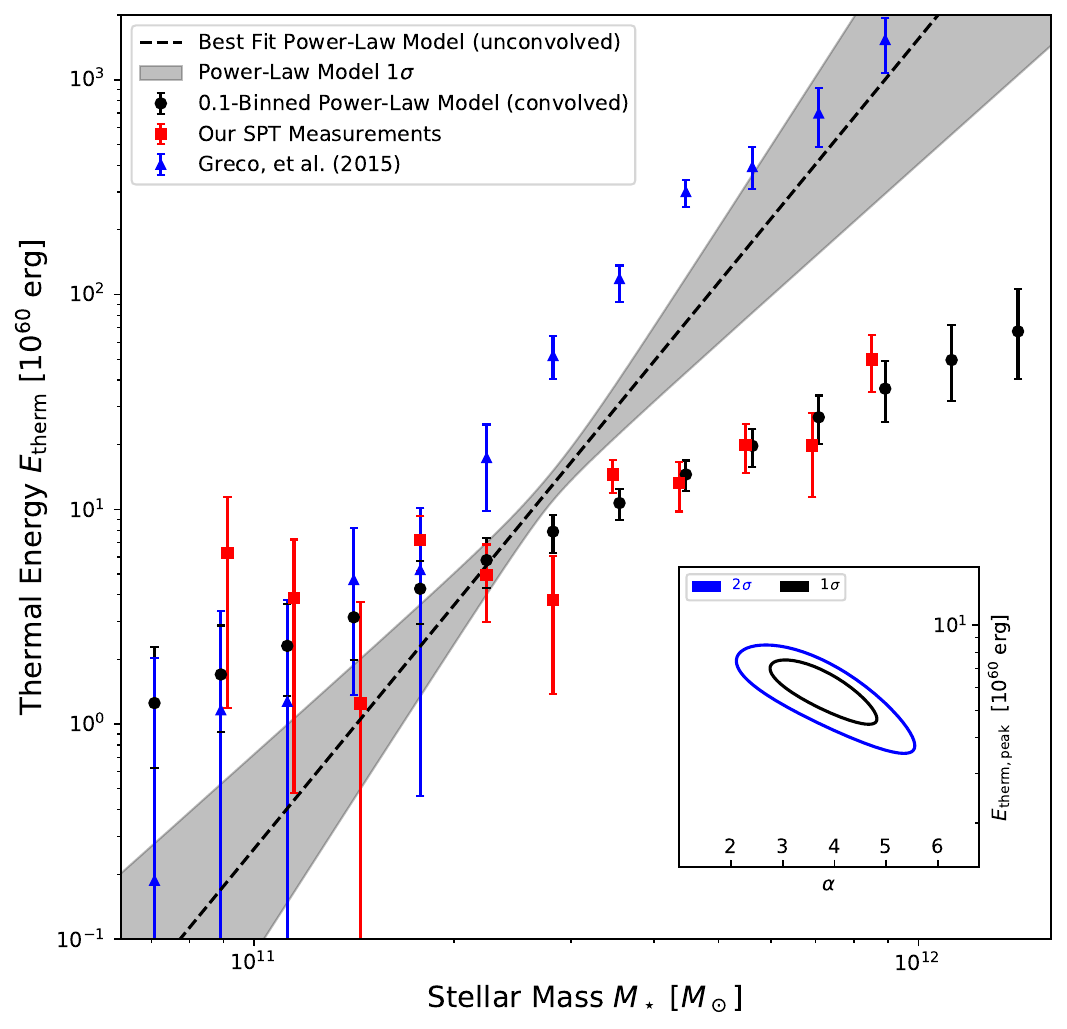}
	\caption{Energy-Stellar Mass plot of the best power-law fit (eq.~\ref{eq:powerlaw}, $E_{\rm therm,peak}=5.98_{-1.00}^{+1.02} \times 10^{60}$ erg, $\alpha=3.77_{-0.74}^{+0.60}$) and shaded $1\sigma$.  Reconvolved thermal energy of 0.1-wide bins from the best fit with $1\sigma$ uncertainty (\textit{black circles}), alongside our original SPT bin measurements (\textit{red squares}) and those converted from \cite{Greco2015} (\textit{blue triangles}). \textit{Inset: } Probability contour for the unconvolved power-law parameters $E_{\rm therm,peak}$ and $\alpha$, with lines at $1\sigma$, and $2\sigma$.}
	\label{fig:powerlaw-fit}
\end{figure*}

\subsection{Correction for Stellar Mass Uncertainty}\label{subsec:masscorrection}

At 0.5 $\leq$ z $\leq$ 1.5 redshifts, our photometrically selected galaxies yield higher uncertainties in stellar mass compared to previous spectroscopic studies \citep{Greco2015}.  To quantify our uncertainties in stellar mass, we performed two Monte Carlo tests. Using the final sample, we first perturbed the photometry using the 1-$\sigma$ photometric errors for each galaxy and band, and  re-ran the SED-fitting code.  This was repeated 100 times for each galaxy, and the standard deviation was computed  for each galaxy. We found the mean uncertainty, due to photometric errors, to be $\sigma_{\log(M_{p})}\simeq0.140$~dex.  Similarly, we repeated the procedure, this time keeping the photometry fixed  but perturbing the photometric redshift using a $5\%$ uncertainty in $1+z$, or $\sigma_{z}\approx0.05(1+z)$. Again, this was repeated 100 times for each galaxy, and we found the mean uncertainty, due to photometric redshift errors, to be $\sigma_{\log(M_{z})}\simeq0.074$~dex. These two numbers were combined in quadrature to give a total estimated uncertainty in stellar mass  of $\sigma_{\log(M)}\simeq0.16$~dex.

This $0.16$~dex uncertainty is large enough to `flatten' our $0.1$~dex stellar mass bin measurements, by shifting a significant number of galaxies near the peak of the mass distribution into wings where they can overwhelm signal from the much smaller galaxy counts at low and high masses.  As an illustration of this effect, Fig.~\ref{fig:EnergyPlots} in Appendix~\ref{Appendix:Greco} shows the results of an additional $0.16$ dex uncertainty applied to the low-redshift galaxies recreated from \cite{Greco2015}.

Our high redshift galaxies also contain this additional artifact from stellar mass uncertainty, but amplified further due to the narrower distribution of our sample.  We correct for this by fitting our ($\log_{10}$) stellar mass distribution to a Gaussian, with a mass peak of $\text{log}_{10}(M_{\star,{\rm peak}}/M_\odot)=11.36 \pm 0.001$ and $\sigma=0.20\pm0.001$.  This allows for easy deconvolution of the $0.16$ dex stellar mass uncertainty, yielding an unconvolved distribution of $\sigma_{\rm uncon.}=\sqrt{(0.20)^2-(0.16)^2}=0.12$.  We assign a simple power-law energy-mass function to the unconvolved distribution following:
\begin{equation}
\label{eq:powerlaw}
	E_{\rm therm}= E_{\rm therm,peak} \left(\frac{M_{\star}}{M_{\rm \star, peak}}\right)^\alpha,
\end{equation}
where $M_{\rm \star,peak} = 2.29 \times 10^{11} M_\odot.$

We assign the energies of this model to the unconvolved ($\sigma=0.12$) $\text{log}_{10}(M_\star/M_\odot)$ distribution and convolve them with our $0.16$ dex uncertainty.  This brings our mass distribution back to the original $\sigma=0.20$ Gaussian fit.  The now-convolved assigned energies can then be placed in similar $0.1$ dex stellar mass bins and fit to our measured SPT values to find $E_{\rm therm,peak}$ and $\alpha$.  Figure~\ref{fig:powerlaw-fit} shows the probability contour of our two-parameter fit (\textit{inset}) and the corresponding unconvolved power-law relation with shaded $1\sigma$, alongside the reconvolved 0.1-wide bin values used in the fit, our original SPT measurements, and thermal energies extracted from \cite{Greco2015}.

This analysis reveals that thermal energy is indeed noticeably flattened at low and high mass bins for our stellar mass uncertainty of $0.16$ dex.  We extract a basic energy-stellar mass relation following eq.~(\ref{eq:powerlaw}) with $E_{\rm therm,peak}= 5.98_{-1.00}^{+1.02} \times 10^{60}$ erg and $\alpha=3.77_{-0.74}^{+0.60}$, indicative of the expected signal after stellar mass uncertainty correction.  This relation also corresponds closely to the lower redshift investigations of \cite{PlanckCollaboration2013,Greco2015}.

\subsection{Implications for AGN Feedback}\label{subsec:Feedback}

While detailed constraints on AGN models are best carried out with comparisons to full numerical simulations, we can nevertheless draw general inferences from our constraints on $E_{\rm therm}.$ The first of these is the remarkable similarity between our $z\approx 1$ results and the  \cite{Greco2015} at $z \approx 0.1.$   It is important to note that these samples were selected by applying slightly different criteria.  In our case, we apply cuts on age~$>1$ Gyr, $SSFR <0.01$ Gyr$^{-1},$ while \cite{Greco2015} selects locally brightest galaxies, defined as brighter than all other sample galaxies within a 1000 km/s redshift difference and projected distance 1.0 Mpc. On the other hand, a wide range of theoretical models suggest a good match between the most massive quiescent galaxies at moderate redshifts and the central galaxies of massive halos in the nearby universe \cite[e.g][]{Moster2013,Schaye2015,Pillepich2018}.

This lack of significant evolution of thermal energy in the CGM around massive galaxies since $z\approx 1$ parallels the lack of significant evolution in the luminosity function of these galaxies \cite[e.g.][]{vanDokkum2010,Muzzin2013}. All in all, this trend is slightly more compatible with models in which AGN feedback is dominated by radio mode contributions.  This is because in such models, gas accretion contributes to CGM heating and radiative losses will contribute to CGM cooling.  Whenever cooling exceeds heating, jets will arise that quickly push the gas up to the constant temperature and entropy at which cooling is inefficient. In quasar models on the other hand, the energy input from feedback occurs once at high redshift, and gas is heated to the point that cooling is extremely inefficient up until today.  In this case, gravitational heating will increase $E_{\rm therm}$ without any significant mechanism to oppose it. However, the particulars of this evolution are highly dependent on the history of galaxy and halo mergers between $0 < z \lesssim 1$.  Hence, it is possible that some types of quasar dominated models may be compatible with our measurements. 

A second major inference is the overall level of feedback. As an estimate of the magnitude of gravitational heating, we can assume that the gas collapses and virializes along with an encompassing spherical dark matter halo, and is heated to the virial temperature $T_{\rm vir}$.  This gives 
\begin{equation}
E_{\rm therm,halo}(M_{13},z)=
1.5 \times 10^{60} \, {\rm erg} \, M_{13}^{5/3} (1+z),
\label{eq:Egravhalo}
\end{equation}
where $M_{13}$ is the mass of the halo in units of $10^{13} M_\odot$ \citep{Spacek2016}. For massive elliptical galaxies, we can convert from halo mass to galaxy stellar mass  using the observed relation between black hole mass and halo circular velocity  \citep{Ferrarese2002}, and the relation between black hole mass and bulge dynamical \citep{Marconi2003}. As shown in \cite{Spacek2016}, this gives 
\begin{equation}
E_{\rm therm,gravity}(M_{\star},z)  \approx  5 \times 10^{60} \, {\rm erg} \, \frac{M_{\star}}{10^{11}M_{\odot}} (1+z)^{-3/2}.
\label{eq:Egrav}
\end{equation}
This is the total thermal energy expected around a galaxy of stellar mass $M_{\star}$ ignoring both radiative cooling and feedback.   For a mean redshift of $\approx 1.1$ and  $M_{\rm \star,peak} = 2.29  \times 10^{11} M_\odot$ this gives $\approx  4 \times 10^{60} \, {\rm erg}.$ 
Note that this estimate has an uncertainty of about a factor of two, which is significantly larger than the uncertainty in our measurements. Nevertheless it is somewhat lower than $E_{\rm therm,peak} = 5.98_{-1.00}^{+1.02} \times 10^{60} \, {\rm erg},$  suggesting the presence of additional non-gravitational heating, particularly as cooling losses are not included in eq.\ (\ref{eq:Egrav}). 

As a simple estimate of heating due to quasar-mode feedback, we  can make use of the model described in \cite{Scannapieco2004}, which gives
\begin{equation}
E_{\rm therm, feedback} (M_{\star},z) \approx  4  \times 10^{60} \, {\rm erg} \,  \epsilon_{k,0.05}  \, \frac{M_{\star}}{10^{11}M_{\odot}}\, (1+z)^{-3/2},
\label{eq:EAGN}
\end{equation}
where $\epsilon_{k,0.05}$ is the fraction of the bolometric luminosity of the quasar associated with an outburst, normalized by a fiducial value of 5\%, which is typical of quasar models  \cite[e.g.][]{Scannapieco2004,Thacker2006,Costa2014}.  For the peak mass and average redshift of our sample, this gives $\approx  \epsilon_{k,0.05} 3 \times 10^{60} \, {\rm erg}$.  Taking $\epsilon_{k,0.05} = 1$ and adding this to contribution from $E_{\rm therm,gravity}$ above gives a total energy of $\approx 7 \times 10^{60} \, {\rm erg}.$ This is on the high side of our measurements, but as it does not account for any energy losses, appears to be a somewhat better match than models without non-gravitational heating. 

Finally, radio mode models are expected to fall somewhere between these two limits, with jets supplying power to roughly balance cooling processes, but never adding a large burst of additional energy of the type, estimated in eq. (\ref{eq:EAGN}).  This would suggest a somewhat better match to the data than pure-gravitational heating models, but again with far too much theoretical uncertainty to draw any definite conclusions.

A third major inference from our measurements comes from the slope of eq.\ (\ref{eq:powerlaw}), which is significantly steeper than in our simple models.  This is most likely due to uncertainties in the halo-mass stellar mass relation, which are particularly large for the most massive $z \approx 1$ galaxies \citep{Wang2013,Lu2015,Moster2018,Kravtsov2018,Behroozi2010,Behroozi2019}.  This represents a major change in the field from only a few years ago, in which tSZ detections at halo masses smaller than galaxy clusters were only marginal, and provided only weak constraints on feedback.  Rather, our measurements, along with other recent constraints \citep{Schaan2020,Amodeo2020}, make it clear that observations are fast outstripping theoretical estimates, and that future close comparisons between measurements and full simulations will yield significant new insights into the history of AGN feedback.

\subsection{Two-Halo Effect}\label{subsec:TwoHalo}

\begin{figure*}[t] 
	\centering\includegraphics[scale=0.5]{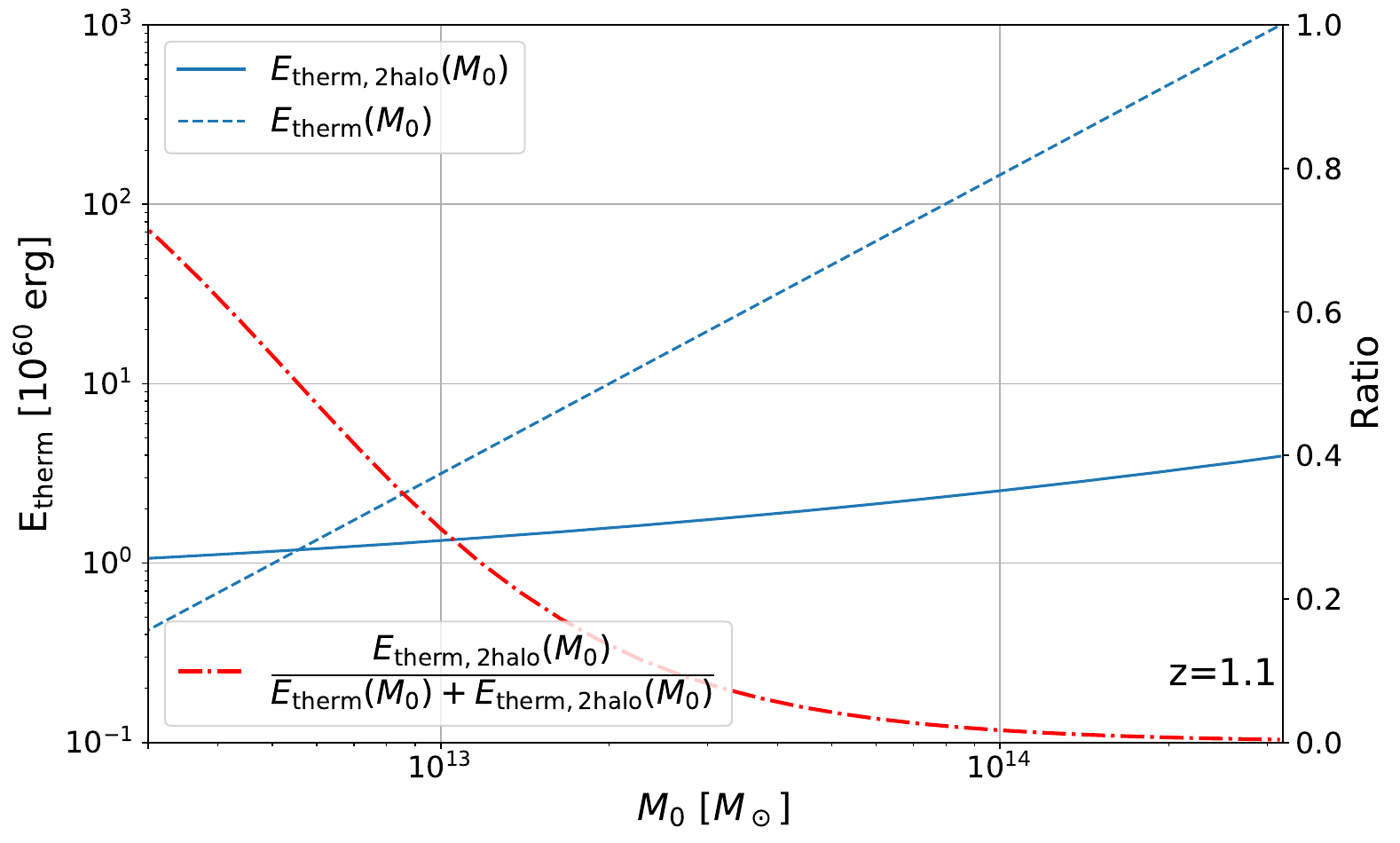}
	\caption{Contribution to the thermal energy from a two-halo term.  The dashed blue line is an estimate of the gravitational thermal energy in the CGM of a galaxy with a given halo mass (x-axis). The solid blue line estimates the total thermal energy in a cylinder (radius = 2 Mpc x len = 400 Mpc) centered and stacked on a galaxy with a given halo mass. This estimate uses the 2-point correlation function to  include galaxies within the cylinder given the mass of the central galaxy, $M_0$ (see details in the text). All data is for z=1.1.  The red dot-dashed line indicates the ratio of the energy in the cylinder to the sum of both curves and uses the y-axis to the right, which shows that the central halo dominates the total energy when its mass exceeds $\approx 10^{13} M_\odot$ or when $E_{\rm therm}$ exceeds $\approx 3 \times 10^{60}$ erg.}
\label{fig:TwoHalo}
\end{figure*}

It is important to recognize that our measurements include the contribution not only from the selected galaxies, but also from the excess of galaxies clustered around them.  This so-called  two-halo contribution is described in detail in \cite{Hill2018}  where it is shown to be significant for lower redshift galaxies and clusters.  To  estimate the impact of this effect on our $z \approx 1$ sample, we can again make use of a simple model for gravitational heating as given by  eq.\ (\ref{eq:Egravhalo}).
This allows us to compute the excess energy due to the neighboring halos within a radius $R$ perpendicular to a central galaxy of halo mass $M_0$ as
\begin{equation}
E_{\rm therm, 2halo} = b(M_0)  \int_{10^{11}\, \rm M_\odot}^{10^{16}\, \rm M_\odot}{\frac{dn(M,z)}{dM}\, E_{\rm therm,halo}(M,z)\, b(M)\, dM}  {\int_{-100 R}^{100 R} 2 \pi {\int_{0}^{R} dr \, r\, \xi(\sqrt{r^2+l^2},z)\,dr}\,dl},
\label{eq:Egravtwohalo}
\end{equation}
where we account for the excess of neighboring halos with masses between  $10^{11} - 10^{16}$ M$_\odot$ within a cylinder of radius $R$ and length $200 R.$ 
In  this expression, $dn(M,z)/dM$ is the number density of dark matter halos per unit mass, $\xi(r,z)$ is the dark matter correlation function, and $b(M) = 1+ (\nu^2-1)/1.69$ (with $\nu \equiv  1.69 \sigma(M,z)^{-1}$)  is bias factor that accounts for mass-dependent differences between the underlying dark matter density field and the distribution of massive dark matter halos \citep{Mo1996}.  

In Fig.~\ref{fig:TwoHalo} we compare the energy in the central halo given by eq.\ (\ref{eq:Egravhalo}) to the two-halo contribution given by eq.\ (\ref{eq:Egravtwohalo}),
over the range of halo masses from $M_0 = 10^{12.5}-10^{14.5}\, M_\odot.$  In this figure, we use the Code for Anisotropies in the Microwave Background \citep{2011ascl.soft02026L} to compute the dark matter correlation function, $\xi(r,z),$ and the rms fluctuations at a redshift $z$ within a sphere containing a mass $M$, $\sigma(M,z),$ and we compute the number density of halos, $dn/dM(M,z),$ according to the standard \cite{Press1974} formula.

While the relationship between galaxy stellar mass and halo mass at $z\approx 1.1$ is uncertain as mentioned above, the fact that the central halo becomes $E_{\rm therm, halo} \gg E_{\rm therm, 2halo}$ when the total thermal energy exceeds $\approx 3 \times 10^{60}$ erg suggests that the majority of our measurements are in the range for which the central halo is dominant. The reason that the two halo term is significantly less important than at $z \approx 0$ is because the dark matter structures have had less time to collapse, meaning that the matter correlation function $\xi$ is significantly smaller than at $z \approx 1$ than it is today.  Thus, although this effect merits careful consideration in comparisons with simulations, it is less likely to complicate the analysis to the degree it does for more local samples \citep[e.g.][]{Schaan2020}.  In addition, although this cannot be computed directly from eq.\ (\ref{eq:Egravtwohalo}), the two halo contribution is likely to become more important on sightlines with large impact parameters from the selected galaxies, and thus it is important to keep in mind when interpreting the tSZ radial profile, a topic to which we turn our attention in the next section.

\subsection{Radial Profile}\label{subsec:RadialProfile}

\begin{figure*}[t]
	\centering\includegraphics[scale=0.65]{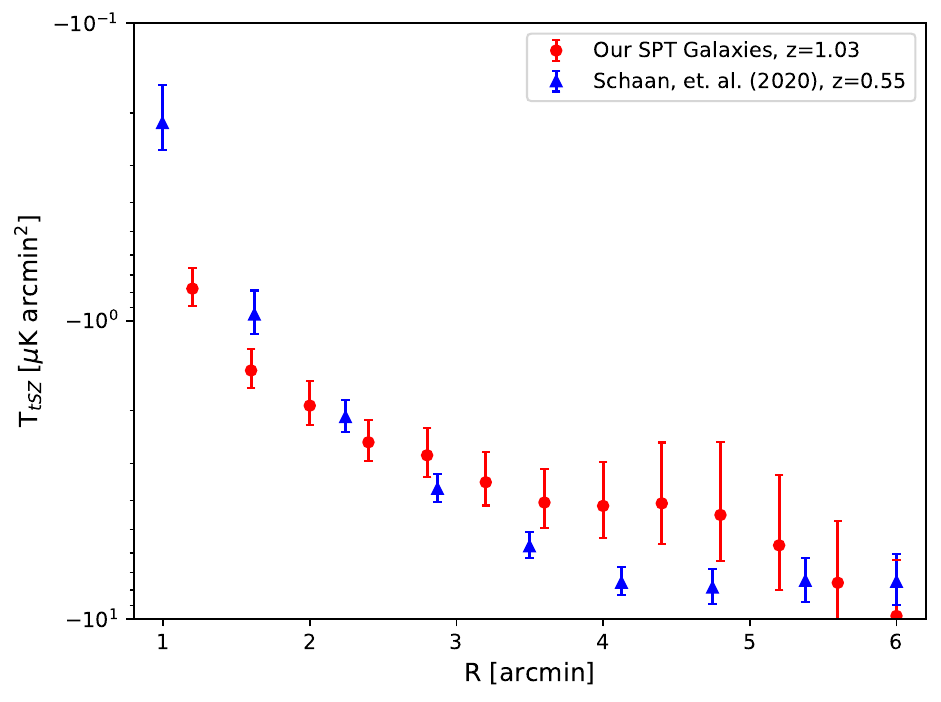}
	\caption{Radial profile of our complete (N=138235) SPT galaxy catalog (\textit{red circles}, $1 \text{ arcmin}=1.01 \text{ comoving Mpc}=0.50 \text{ proper Mpc } @ z=1.03$), alongside the recent profile from \cite{Schaan2020} (\textit{blue triangles}, $1 \text{ arcmin}=0.61 \text{ comoving Mpc}=0.39 \text{ proper Mpc } @ z=0.55$).  The y-axis is in units of integrated CMB temperature at 150 GHz ($\mu$K arcmin$^2$).  The two profiles are similar even with their difference in redshift.}
	\label{fig:RadialProfile}
\end{figure*}

The radial profile of the tSZ signal has been recently studied at lower redshifts with the latest ACT release at similar angular resolution  \citep{Schaan2020}.  To compare against this study, we apply the same procedure as in \cite{Schaan2020} to compute this profile: summing within a cap or top-hat shape of radius $R$, and subtracting the neighboring pixels out to $\sqrt{2}R$ to remove the surrounding background offset such as from the primary CMB.  In this case we abstain from the high-pass filtering done in our main stellar mass bin analysis.

The profile of our SPT galaxies is shown in Fig.~\ref{fig:RadialProfile}, where it is compared to the CMASS sample in  \cite{Schaan2020} which has a mean redshift of $z=0.55$ and mean $\log_{10}(M_\star/M_\odot)$ in linear units is $\approx 2 \times 10^{11} M_\odot.$  As a preliminary analysis, we stack all N=138,325 SPT galaxies, and apply the same style aperture from $R=1.2-6.2$ arcmin radius, at each band.  We then use our two-component fit of eq.~(\ref{eq:SZFit}) using the median redshift of $z=1.03$, and dust parameters T $=20\pm5$K and $\beta=1.75\pm0.25$ applied as before.  The final tSZ values are converted to integrated CMB temperature with respect to 150 GHz, as in \cite{Schaan2020}.  The two tSZ profiles have very similar overall shapes.  Our profile however is slightly less defined with larger uncertainty, due to being at higher redshift and a smaller sample size.  A more detailed comparison, such as by mass-binning and fitting to profile models, is left for future work.

\section{Discussion}\label{sec:Discussion}

From $z \approx 1$ to $z \approx 0,$  dark matter halos continue to merge and accrete material,  but
the growth of massive galaxies over this redshift range is minimal, with  $\log_{10}(M_\star/M_\odot)=11.15$ galaxies growing in mass by less than $0.2$ dex \citep{Muzzin2013}.  To explain this surprising trend, theoretical models have been forced to invoke significant energy input from AGN, which heats the medium surrounding massive galaxies to temperatures high enough to prevent it from cooling and forming further generations of stars.  On the other hand,  such AGN feedback is largely unconstrained by observations.  

Here we make use of measurements of the tSZ effect to derive direct constraints that can be applied to such models.  As the total tSZ distortion along a given sightline is proportional to the line-of-sight integral of the pressure, the total signal summed over the area of sky around any object  is proportional to the volume integral of the pressure, or the total thermal energy. This means that by summing the Compton distortions over the patches of sky around galaxies, we can directly measure the thermal energy of the CGM surrounding them.

We apply this technique to constrain the signal of 138,235 $z\approx 1$ galaxies, selected from the DES and WISE surveys. Data from the SPT at 95, 150, and 220 GHz were stacked around the galaxies, spatially filtered to separate the signal from primary CMB fluctuations, and fit with a gray-body model to remove the dust contribution, which is detected with a signal to noise ratio of $9.8\sigma$. The resulting tSZ around these galaxies is detected with an overall signal to noise ratio of $10.1\sigma$, which is large enough to allow us to partition the galaxies into 0.1 dex stellar mass bins from $M_{\star} = 10^{10.9} M_\odot$ - $10^{12} M_\odot$, which have corresponding tSZ detections of up to $5.6\sigma$.  We also observe significantly more dust at these frequencies than previous low redshift studies \citep{PlanckCollaboration2013,Greco2015}, and a noticeable increase in dust signal with larger mass bins. 

As the stellar mass distribution of our selected galaxies is highly peaked at $M_{\star, {\rm peak}} = 2.3 \times 10^{11} M_\odot$, the 0.16 dex uncertainty in our photometric fits to the masses is large enough to `flatten' our measurements, shifting a significant number of galaxies near the distribution peak into wings where they can overwhelm signal from the much smaller galaxy counts at low and high masses.
To correct for this effect, we carry out a two parameter fit for the unconvolved energy-mass relation that best fits our data, of the form $E_{\rm therm} = E_{\rm therm,peak} \left(M_\star/M_{\star,{\rm peak}} \right)^\alpha$.   In this case, we find an amplitude at the mass peak of $E_{\rm therm,peak}= 5.98_{-1.00}^{+1.02} \times 10^{60}$ erg and a slope of $\alpha=3.77_{-0.74}^{+0.60}$.  This, however, would not take into account any inherent biases in our stellar mass uncertainty, such as a potentially smaller uncertainty for our brighter, more massive galaxies.  

This aligns well with previous $z \approx 0$ studies, indicating a good match between the thermal energy of the CGM surrounding the most massive quiescent galaxies at moderate redshifts, and the central galaxies of massive halos in the nearby universe. When compared to theoretical models, our energy-mass relation best corresponds to moderate radio mode feedback.  Purely gravitational heating predictions are slightly lower than the final results, while quasar-mode AGN feedback models are slightly higher.  However, all of these models have wide enough uncertainties that definitive conclusions are difficult to be drawn from them.  This means that observations are no longer in the regime of marginal detections, and are quickly outstripping the capabilities of theoretical estimates.  This further highlights the need for improved theoretical models to keep pace with the ever-increasing observational capabilities.

The limitations in our observational analysis primarily stem from our ability to accurately fit and remove the large dust component such as seen in the galaxy stacks in Fig.~\ref{fig:AllStackZoomed} and the dust fit in Fig.~\ref{fig:FitPlots}.  The dust fit amplitude is set mainly by the 220 GHz band, and we use a spectral emissivity index $\beta$ and dust temperature $\text{T}_{\rm dust}$ with large uncertainty due to our inability to fit them with only three frequency bands.  The addition of a similar resolution survey at far-infrared frequencies would provide a better fit of all thermal dust parameters.

We also recognize there is a likely two-halo contribution term for these stacking type measurements, which was shown to be large in \cite{Hill2018} for lower redshift galaxies and clusters.  While simple models such as presented in \S\ref{subsec:TwoHalo} suggest that the contribution of this effect to the total SZ signal is small for the majority of our $z \approx 1$ galaxies, the exact significance of the two-halo contribution to our measurements of the radial profile are yet to be determined.  

An initial radial tSZ profile of our entire galaxy sample in \S\ref{subsec:RadialProfile} highlights the similarities with lower redshift studies \citep{Schaan2020}.  Additional investigation will be needed to refine and compare to profile models, including the impact of additional sources such as the aforementioned two-halo term.   

Similar resolution surveys across more of the sky, such as planned with the next generation Atacama Cosmology Telescope instrument \citep[Advanced ACTPol][]{Koopman2018}, will enable a larger sampling area.  Likewise, advances in spatial resolution, such as will be possible using the TolTEC camera being built for the 50-meter Large Millimeter-wave Telescope \citep{Bryan2018},  will allow for a cleaner separation between the tSZ signal, which comes primarily from the CGM, and the dust signal, which comes primarily from the underlying galaxy.  Such developments promise to yield  dramatic new insights into the physical processes that shaped the most massive galaxies in the universe.

\acknowledgments

We thank Alexander Van Engelen for helpful discussions and thank the referee for detailed comments,  which greatly improved this work. We acknowledge Research Computing at Arizona State University for providing HPC resources that have contributed to the research results reported within this paper.  The galaxy data used here was from DES and WISE, while the maps are publicly obtained from SPT.  This publication makes use of data products from the Wide-field Infrared Survey Explorer, which is a joint project of the University of California, Los Angeles, and the Jet Propulsion Laboratory/California Institute of Technology, and NEOWISE, which is a project of the Jet Propulsion Laboratory/California Institute of Technology. WISE and NEOWISE are funded by the National Aeronautics and Space Administration.  The manipulation of the maps was aided by the \textit{Healpy} python module for Healpix operations.  ES was supported by NSF grant AST-1715876.

\appendix

\begin{figure*}[t]
	\centering\includegraphics[scale=0.7]{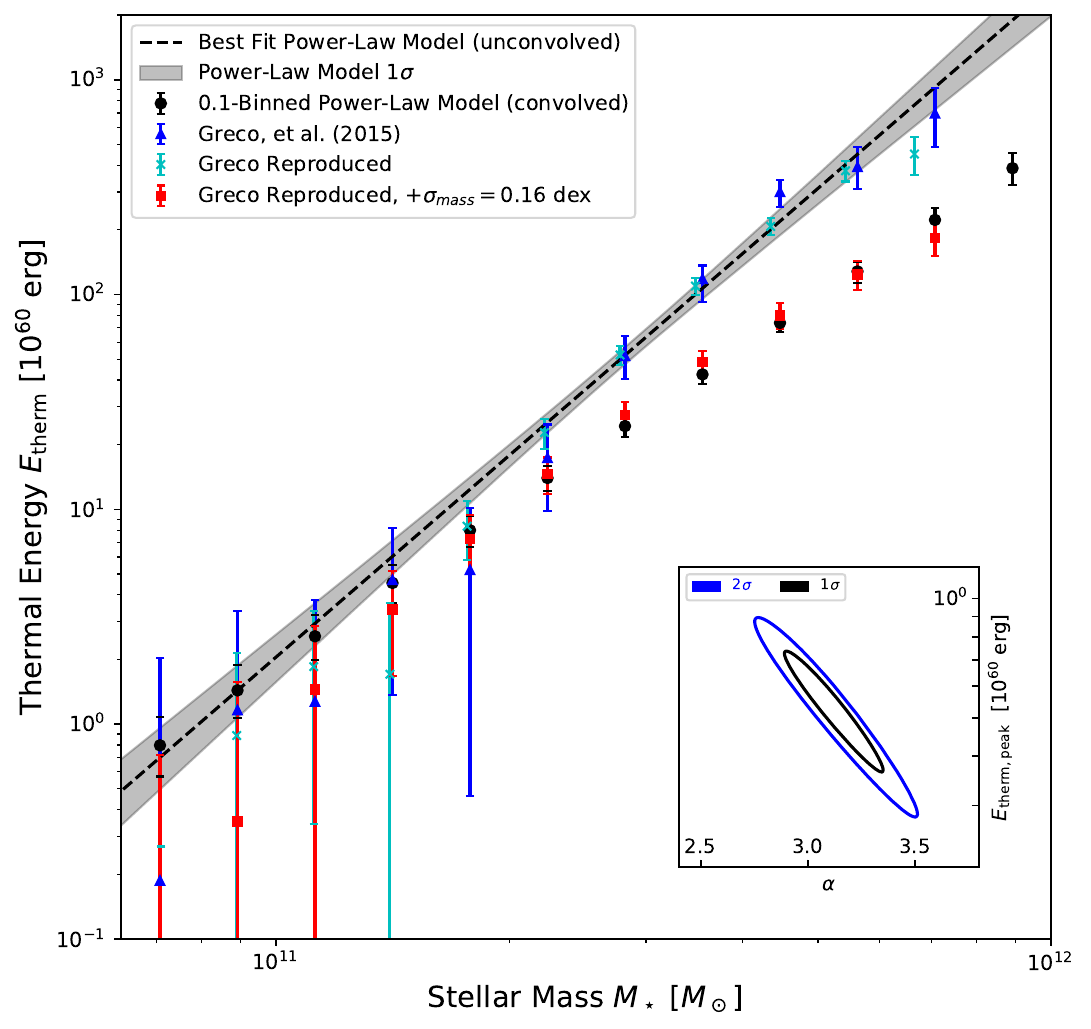}
	\caption{Low-z galaxies observed in \cite{Greco2015} converted to thermal energy (\textit{blue triangles}).  We also recreated the low-z galaxy sample as described in \cite{Greco2015,PlanckCollaboration2013} and obtained similar results (\textit{cyan crosses}).  Finally, a major concern with our high-z SPT galaxies is the larger uncertainty in mass ($\sigma_m=0.16$ dex).  Using the recreated low-z galaxies, an additional $\sigma_m=0.16$ dex uncertainty in mass shows diluted signal in the higher stellar mass bins, indicating our high-z galaxies contain this effect (\textit{red squares}). Error bars represent $1\sigma$ uncertainties.  We also then show our energy-mass function of eq.~(\ref{eq:powerlaw}) applied to the flattened bins, producing a slope that aligns well with the original values.  The flattened power-law bins used in the fit are shown as well (\textit{black circles}).}
	\label{fig:EnergyPlots}
\end{figure*}

 \section{Reproduction of Previous Low Redshift Results}\label{Appendix:Greco}
To ensure there were no unknown errors in the stacking procedure and code, we reconstructed the locally brightest galaxy catalog used in the studies of \cite{PlanckCollaboration2013,Greco2015} and stacked them with the Planck Modified Internal Linear Combination Algorithm (MILCA) Compton-$y$ map.

The catalog was selected in identical fashion as was done in \cite{PlanckCollaboration2013,Greco2015}.  Starting from the spectroscopic New York University Value Added Galaxy Catalog\footnote{\url{https://sdss.physics.nyu.edu/vagc/}} \citep{Blanton2005}, which uses the Sloan Digital Sky Survey (SDSS) DR7, $r<17.7$ magnitude and $z>0.03$ cuts were first applied.  Then any smaller neighbors of the catalog were removed by selecting only galaxies brighter than all other sample galaxies within a 1000 km/s redshift difference and a projected distance of 1.0 Mpc.  This method was then repeated with a separate photometric catalog, photoz2 \citep{Cunha2009}.  After a final removal of any galaxies within Planck point sources and 40\% galactic masks, 243364 galaxies were obtained.  This is marginally lower than that reported in \cite{PlanckCollaboration2013}, likely from a slightly stricter removal of flagged galaxies or updated masks.  Since spectroscopic and low redshift, these have a lower stellar mass uncertainty of $\approx$0.1 dex \citep{Blanton2007} than our photometric and high-z catalog ($\approx$0.16 dex).  We bin the low-z galaxies in similar $\log_{10}(M_\star/M_\odot)$ 0.1-wide mass bins.

The map used for stacking is a component-separated Modified Internal Linear Combination Algorithm (MILCA) from Planck DR2 \footnote{\url{https://irsa.ipac.caltech.edu/data/Planck/release\_2/all-sky-maps/ysz\_index.html}} \citep{PCXXII}.  This map has units of micro Compton-$y$ and a FWHM of 10 arcmin.  For simplicity, we use an R $=10$ arcmin radius top-hat aperture and mean-subtract with an annulus from R to R$+$FWHM.  The average equal-weighted signals from mass bins of $10.0<\log_{10}(M_\star/M_\odot)<11.9$ are shown in Fig.~\ref{fig:EnergyPlots}, converted to thermal energy as in eq.~(\ref{eq:EthrmT}).  Even without simulations for inverse-variance weighting and variable aperture radii, our results compare well with those of \cite{Greco2015} also plotted in Fig.~\ref{fig:EnergyPlots}.

To determine the effect of our larger $\approx$0.16 dex mass uncertainty in our original high-z catalog, we apply a 0.16 dex Gaussian uncertainty to all low-z galaxy masses.  By doing a new weighted mass binning according to the likelihood of each galaxy to be in each respective bin, we calculate the new average thermal energies with the 0.16 dex mass uncertainty (also shown in Fig.~\ref{fig:EnergyPlots}).  These values reveal a dimming of the signal at high masses, due to contamination from low mass galaxies shifted higher from the uncertainty.  This is likely to be similar for our high-z galaxies, and provides a challenge to be accurately corrected.  Our method aimed at correcting for this (in \S\ref{subsec:masscorrection} and eq.~\ref{eq:powerlaw}) can similarly be applied to these purposely dimmed values to obtain $E_{\rm therm,peak}=0.53_{-0.11}^{+0.14}$ erg and $\alpha=3.12_{-0.19}^{+0.12}$ about a peak mass of $\text{log}_{10}(M_{\star,{\rm peak}}/M_\odot)=11.8$.  As shown in Fig.~\ref{fig:EnergyPlots}, this aligns well with the expected original results and the slope $\alpha$ is near the $3.77_{-0.74}^{+0.60}$ extracted from our SPT sample in \S\ref{subsec:masscorrection}.

Later studies \citep{Hill2018} have also shown a residual two-halo effect occurs for this low redshift catalog, more so evident at the low masses.  In some instances, the two-halo contribution was shown to dominate, but further investigation will be needed to translate this to our higher redshift galaxies.

\section{Impact of Aperture Size}\label{Appendix:Aperture}
The aperture shape and size used in the integrated sums at each frequency has a large influence on the final two-component fit of eq.~\ref{eq:SZFit}.  It needs to be large enough to encompass most of the signal from our galaxies without being too large to introduce extra noise and nearby source contamination.  

Table~\ref{tab:appendix_apertures} shows the $\text{log}_{10}(M_\star/M_\odot)$ 0.1-wide bin Compton-$y$ fits, alongside the final stellar mass correction power-law parameters $E_{therm,peak}$ and $\alpha$ for three other potential aperture choices.  A 3.0 arcmin FWHM Gaussian aperture is chosen as a similar comparison to our main 2.0 arcmin top-hat used.  Two larger apertures: a 4.0 arcmin top-hat, and a 6.0 arcmin Gaussian are also selected to highlight the additional noise they incur.
\begin{table*}[ht]
	\begin{center}
		\def\arraystretch{1.2}
		\begin{tabular}{|c|c|c|c|c|c|}
			\hline
			& $\text{log}_{10}(M_\star/M_\odot)$ & $2.0$ arcmin & $4.0$ arcmin & $3.0$ arcmin & $6.0$ arcmin \\
		       &   &  Top-Hat & Top-Hat & Gaussian &   Gaussian\\
			\hline 
			\multirow{11}{*}{\rotatebox[origin=c]{90}{ $\int y(\vec{\theta}) d\vec{\theta}$  [$10^{-6}\text{ arcmin}^2$]}}
			& $10.9-11.0$ & $0.78_{-0.63}^{+0.63}$  & $1.23_{-1.35}^{+1.35}$ & $0.72_{-0.61}^{+0.61}$ & $0.90_{-1.02}^{+1.02}$ \\
			& $11.0-11.1$ &  $0.47_{-0.42}^{+0.41}$ & $0.79_{-0.87}^{+0.87}$ & $0.55_{-0.40}^{+0.40}$ & $0.61_{-0.66}^{+0.66}$ \\
			& $11.1-11.2$ & $0.15_{-0.30}^{+0.30}$ & $0.15_{-0.63}^{+0.63}$ & $0.14_{-0.29}^{+0.29}$ & $0.31_{-0.48}^{+0.48}$ \\
			& $11.2-11.3$ & $0.87_{-0.25}^{+0.25}$ & $0.79_{-0.52}^{+0.52}$ & $0.81_{-0.24}^{+0.24}$ & $0.91_{-0.39}^{+0.39}$ \\
			& $11.3-11.4$ & $0.59_{-0.23}^{+0.23}$ & $1.10_{-0.48}^{+0.48}$ & $0.59_{-0.22}^{+0.22}$ & $0.91_{-0.36}^{+0.36}$ \\
			& $11.4-11.5$ & $0.44_{-0.28}^{+0.27}$ & $0.00_{-0.53}^{+0.52}$ & $0.39_{-0.27}^{+0.25}$ & $0.06_{-0.41}^{+0.39}$ \\
			& $11.5-11.6$ & $1.69_{-0.30}^{+0.29}$ & $2.42_{-0.61}^{+0.60}$ & $1.63_{-0.29}^{+0.28}$ & $2.12_{-0.46}^{+0.46}$ \\
			& $11.6-11.7$ &  $1.52_{-0.40}^{+0.39}$ & $1.76_{-0.82}^{+0.81}$ & $1.38_{-0.39}^{+0.38}$ & $1.65_{-0.62}^{+0.62}$ \\
			& $11.7-11.8$ & $2.27_{-0.59}^{+0.57}$ & $2.90_{-1.19}^{+1.18}$ & $2.20_{-0.57}^{+0.56}$ & $2.42_{-0.91}^{+0.90}$ \\
			& $11.8-11.9$ & $2.23_{-0.95}^{+0.94}$ & $0.32_{-2.02}^{+1.99}$ & $2.06_{-0.92}^{+0.91}$ & $1.12_{-1.53}^{+1.50}$ \\
			& $11.9-12.0$ &  $5.57_{-1.65}^{+1.65}$ & $3.78_{-3.47}^{+3.47}$ & $4.91_{-1.60}^{+1.59}$ & $3.45_{-2.63}^{+2.63}$ \\
				\hline \hline
			\multicolumn{2}{|c|}{$E_{\rm therm,peak}$ [$10^{60}$ erg]} & $5.98_{-1.00}^{+1.02}$ & $7.62_{-2.06}^{+1.83}$ & $5.76_{-0.96}^{+0.99}$ & $7.17_{-1.55}^{+1.44}$ \\ \hline
			\multicolumn{2}{|c|}{$\alpha$} & $3.77_{-0.74}^{+0.60}$ & $2.93_{-1.33}^{+0.94}$ & $3.65_{-0.76}^{+0.60}$ & $2.88_{-1.08}^{+0.82}$ \\  
			\hline  
		\end{tabular}
	\end{center}
	\caption{\small Results for alternate aperture shapes and sizes: the fiducial 2.0 arcmin radius top-hat, a larger 4.0 arcmin radius top-hat, a 3.0 arcmin FWHM Gaussian, and a 6.0 arcmin FWHM Gaussian.  Listed are their respective Compton-$y$ parameter fits found from eq.~\ref{eq:SZFit}, and the final two rows showing the stellar mass uncertainty correction power-law fit parameters $E_{therm,peak}$ and $\alpha$.  As expected, the additional noise from the larger apertures results in worse fits.  The 3.0 Gaussian aperture behaves similar to that of our 2.0 arcmin top-hat chosen for the main results.}
\label{tab:appendix_apertures}
\end{table*}

\vspace{0.4in}
\bibliographystyle{apj}

\bibliography{SPT_sz_stack}

\end{document}